\documentclass[10pt, a4paper, twocolumn]{IEEEtran}

\usepackage{amsmath,epsfig,amssymb,verbatim,amsopn,multirow}
\usepackage{balance}
\usepackage[usenames,dvipsnames]{color}
\usepackage[all]{xy}  %%% used to make block diagram
\usepackage{url}
\usepackage{amsfonts}
\usepackage{amssymb}
\usepackage{epsfig}
\usepackage{slashbox}
\usepackage{bm}  %% to access bold math symbols
\usepackage{amsmath,amssymb, amstext}
\usepackage{graphicx}
\usepackage{epstopdf}
\usepackage{hyperref}
\usepackage[english]{babel}
\usepackage{graphicx}
\usepackage{caption}
\usepackage{enumerate}
\usepackage{epsfig}
\usepackage{bm}
\usepackage{amsmath,amssymb, amstext}
\usepackage{graphicx}
\usepackage{epstopdf}
\usepackage{amsmath,epsfig,amssymb,verbatim,amsopn,cite,subfigure,multirow}
\usepackage{thmtools}
\usepackage{mathtools}
\usepackage{balance}
\usepackage[usenames,dvipsnames]{color}
\usepackage[all]{xy}
\usepackage{url}
\usepackage{amsfonts}
\usepackage{amssymb}
\usepackage{color}
\usepackage{bm}
\usepackage{epsfig}
\usepackage{amsmath,amssymb, amstext}
\usepackage{graphicx}
\usepackage{epstopdf}
\usepackage{algorithm}
\usepackage{algorithmic}
\usepackage{subfigure}
\usepackage[subfigure]{tocloft}
\usepackage[mathscr]{euscript}
\usepackage{multirow}
\usepackage{cleveref}
\usepackage[font=small,labelfont=bf,labelsep=space]{caption}
\newtheorem{theorem}{Theorem}

\newtheorem{definition}{Definition}

\newtheorem{remark}{Remark}

\newenvironment{proof}%
  {\proof}{\proofend}

\newcounter{mytempeqcounter}

\renewcommand\thesubsection{\thesection.\arabic{subsection}}
\renewcommand{\figurename}{Fig.}
\newcommand{\superscript}[1]{\ensuremath{^{\textrm{#1}}}}

\definecolor{light-gray}{gray}{0.65}

\begin{document}
%%%%%%%%%%%%%%%%%%%%%%%%%%%%%%%%%%%%%%%%%%%%%%%%%%%% TITLE %%%%%%%%%%%%%%%%%%%%%%%%%%%%%%%%%%%%%%%%%%
\title{{\Huge{Multiple Access Channel with Common Message and Secrecy constraint}}}

\author{\IEEEauthorblockN{ Hassan Zivari-Fard\superscript{\dag,}\superscript{\dag\dag}, Bahareh Akhbari\superscript{\dag\dag}, Mahmoud Ahmadian-Attari\superscript{\dag\dag}, Mohammad Reza Aref\superscript{\dag}}

\IEEEauthorblockA{
\superscript{\dag}Information Systems and Security Lab (ISSL), Sharif University of Technology, Tehran, Iran\\
Email: {{hassan\_zivari}@ee.kntu.ac.ir, {aref}@sharif.edu}\\
\superscript{\dag\dag}Department of ECE, K. N. Toosi University of Technology, Tehran, Iran\\
Email: {\{akhbari, mahmoud\}@eetd.kntu.ac.ir}}}
\maketitle
\date{}
%%%%%%%%%%%%%%%%%%%%%%%%%%%%%%%%%%%%%%%%%%%%%%%%%%%% ABSTRACT %%%%%%%%%%%%%%%%%%%%%%%%%%%%%%%%%%%%%%%
%\footnote{This work was partially supported by Iran Telecom Research Center under contract no. 17175/500.}
\begin{abstract}
In this paper, we study the problem of secret communication over a multiple-access channel with a common message. Here, we assume that two transmitters have confidential messages, which must be kept secret from the wiretapper (the second receiver), and both of them have access to a common message which can be decoded by the two receivers. We call this setting as Multiple-Access Wiretap Channel with Common message (MAWC-CM). For this setting, we derive general inner and outer bounds on the secrecy capacity region for the discrete memoryless case and show that these bounds meet each other for a special case called the switch channel. As well, for a Gaussian version of MAWC-CM, we derive inner and outer bounds on the secrecy capacity region.
Providing numerical results for the Gaussian case, we illustrate the comparison between the derived achievable rate region and the outer bound for the considered model and the capacity region of compound multiple access channel.
\end{abstract}
\IEEEpeerreviewmaketitle
%\vspace{-4mm}
%%%%%%%%%%%%%%%%%%%%%%%%%%%%%%%%%%%%%%%%%%%%%%%%%%% INTRODUCTION %%%%%%%%%%%%%%%%%%%%%%%%%%%%%%%%%%%%
\section{Introduction}
%\fontsize{10}{11}
%\selectfont
In a seminal work, the wire-tap channel was introduced by Wyner \cite{Wyner}, where a sender wishes to communicate a message to a receiver while keeping the message secret from an eavesdropper. He established the secrecy capacity for a single-user degraded wire-tap channel. Later, Csisz\'{a}r and K\"{o}rner extended the wire-tap channel to a more generalized model called the broadcast channel with confidential messages \cite{CsiszarKorner} and computed its secrecy capacity.

The problem of secret communication over multi-user channels has recently attracted remarkable attention \cite{cooperation, LiangPoor, Yassaee, ISIT2012, GMAWC, Maric, GMAWCJamming, MAWCwithconference, BlochBarros, liumaric, IETsaleh}. In \cite{cooperation, LiangPoor}, Multiple Access Channel (MAC) with generalized feedback has been considered, where in \cite{cooperation} the encoders do not need to keep their messages secret from each other, but their messages should be kept secret from an external eavesdropper. Whereas in \cite{LiangPoor} each user views the other user as an eavesdropper and wishes to keep its confidential information as secret as possible from the other user.

The Multiple Access Wire-tap Channel (MAWC) (i.e. multiple access channel with an external eavesdropper) under strong secrecy criterion has been studied in \cite{Yassaee}. In \cite{ISIT2012}, MAWC has been studied assuming that there exists a common message while the eavesdropper is \emph{unable} to decode it. For this model an achievable rate region for discrete memoryless case under the strong secrecy criterion has been derived.
A degraded Gaussian MAWC, in which the eavesdropper receives a degraded version of the legitimate receiver's signal, has been studied in \cite{GMAWC} and an achievable rate region for this setting has been established. In \cite{GMAWCJamming} general Gaussian MAWC has been considered such that an achievable rate region has been derived. The problem of lossy source transmission over a MAWC was considered in \cite{JSsaleh}.

The influence of partial encoder cooperation on the secrecy capacity of the MAWC has been studied in \cite{MAWCwithconference}. In their considered setting, two encoders that are connected by two communication links with finite capacities wish to send secret messages to the common intended decoder in the presence of an eavesdropper. In their model the transmitters do not have any common messages. 
The compound MAC (two-transmitter/two-receiver MAC) with conferencing links between both encoders and decoders without any secrecy constraint has been studied in \cite{simeone}. In \cite{ICC2014}, we have considered compound MAC with confidential messages so that the first transmitter's private message is confidential and are only decoded by the first receiver, and kept secret from the second receiver, while the common message and the private message of the second transmitter are decoded by both receivers.

In this paper, we investigate secrecy constraints in a multiple access channel with a common message. We call our model as Multiple-Access Wiretap Channel with Common Message (MAWC-CM). To interpret this model, it can be noted that in wireless networks there may be a scenario in which the users may have a \emph{common message} which can be decoded by \emph{all users} in addition to the confidential information that wish to be kept secret from illegal users. Motivated by this scenario, we consider MAWC-CM as a building block of this setting. In this model, each transmitter sends its own private message while both of them have a common message. Both of the transmitter's private messages ($W_1$ and $W_2$) are confidential and are only decoded by the first receiver and kept secret from the second receiver. The common message $W_0$ is decoded by both receivers (see Fig.~\ref{figdiag}). For this model, we derive single-letter inner and outer bounds on the secrecy capacity region. We also study a switch channel which is a special case of our model and show that the derived inner and outer bounds meet each other for this case. We also consider Gaussian MAWC-CM and derive inner and outer bounds on its secrecy capacity region. Providing some numerical examples for Gaussian MAWC-CM, we compare the derived achievable rate region and outer bound for the Gaussian case with each other and also with the capacity region of the Gaussian compound MAC. The considered examples illustrate the impact of noise power and secrecy constraints on the rate regions. We show that there are scenarios for which the secret transmissions may increase achievable rate region in compare with the case that requires the second receiver to decode the private messages.

The rest of this paper is organized as follows. In Section~\ref{notationssec}, the notations and the system model are described. In Section~\ref{discreteinneroutersec}, outer and inner bounds on the secrecy capacity region of discrete memoryless MAWC-CM are established and it is shown that these bounds meet each other for the switch channel model. An achievable secrecy rate region and an outer bound on the secrecy capacity region of Gaussian MAWC-CM are derived in Section~\ref{gaussiansec}. Finally, Section~\ref{conclusions} concludes the paper.
%%%%%%%%%%%%%%%%%%%%%%%%%%%%%%%%%%%%%%%%%%%%%%%%% FIGURE 1 %%%%%%%%%%%%%%%%%%%%%%%%%%%%%%%%%%%
\begin{figure}%*}[ht]
\noindent
\makeatletter%
\if@twocolumn%
\centering
\includegraphics[width=8.50cm]{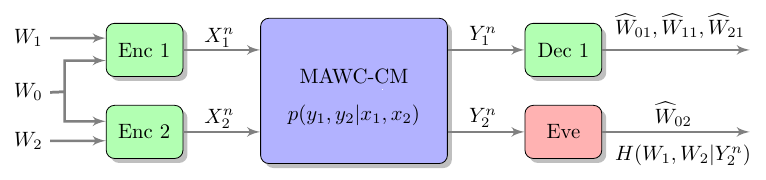}
\caption{Multiple Access Wire-tap Channel with Common Message (MAWC-CM)}
\setlength{\textfloatsep}{10pt plus 1.0pt minus 2.0pt}
\label{figdiag}
\else% \@twocolumnfalse
\centering
\includegraphics[width=12.50cm]{MAWC-CM31.pdf}
\caption{Multiple Access Wire-tap Channel with Common Message (MAWC-CM)}
\setlength{\textfloatsep}{10pt plus 1.0pt minus 2.0pt}
\label{figdiag}
\fi
\makeatother
\vspace{-0.8cm}
\end{figure}%*}
%\vspace{-3cm}
%%%%%%%%%%%%%%%%%%%%%%%%%%%%%%%%%%%%%%%%%%%%%%%%% SYSTEM MODEL %%%%%%%%%%%%%%%%%%%%%%%%%%%%%%
\section{Notations and System Model}\label{notationssec}
In this paper, the random variables are represented by capital letters e.g., $X$, the realizations of random variables are represented by lower case letters e.g., $x$ and their alphabets are represented by ${\cal X}$. The ${\cal T}_\varepsilon ^n(P_{XY})$ indicates the set of $\varepsilon -$strongly jointly typical sequences \cite{BlochBarros} of length $n$, on joint distribution $P_{X,Y}$. $X_i^n$ indicates vector $(X_{i,1},X_{i,2},\ldots,X_{i,n})$, and $X_{i,j}^k$ indicates vector $(X_{i,j},X_{i,j+1},\ldots,X_{i,k})$. The cardinality of $U$ is denoted by $|\cal U|$.

\begin{definition}\label{firstdefi} Consider a discrete memoryless MAWC-CM $({{\cal X}_1},{{\cal X}_2},p({y_1},{y_2}|{x_1},{x_2}),{{\cal Y}_1},{{\cal Y}_2})$ where ${{\cal X}_1}$ and ${{\cal X}_2}$ are the finite input alphabets of transmitters, ${{\cal Y}_1}$ and ${{\cal Y}_2}$ are the channel output alphabets of receiver~1 and receiver~2, respectively (Fig.~\ref{figdiag}) and $p({y_1},{y_2}|{x_1},{x_2})$ is the channel transition probability distribution.
\end{definition}
\begin{definition}\label{seconddefi} A  $({2^{nR_0}},{2^{nR_1}},{2^{nR_2}},n)$ code for the MAWC-CM (Fig.~\ref{figdiag}) consists of the followings:
i) Three message sets ${\mathcal{W}}_u = \{1,...,{2^{nR_u}}\}$ for $u=0,1,2$ where independent messages $W_0$, $W_1$ and $W_2$ are uniformly distributed over respective sets.
ii) Two stochastic encoders ${g_k}$, $k = 1,2$, for transmitter $k$ that are specified by $g_k:{{\cal W}_0}\times{{\cal W}_k}\rightarrow {{\cal X}_k^n}$ for $k=1,2$.
iii) Two decoding functions $\phi :{{\cal Y}_{1}^n} \to {{\cal W}_0} \times {{\cal W}_1} \times {{\cal W}_2}$ and $\rho :{{\cal Y}_{2}^n} \to {{\cal W}_0}$. The first decoder is at the legitimate receiver and assigns $({\widehat W_{01}},{\widehat W_1},{\widehat W_2}) \in {{\cal W}_0} \times {{\cal W}_1} \times {{\cal W}_2}$ to each received sequence $y_1^n$. The second decoder assigns an estimate ${\widehat W_{02}} \in {{\cal W}_0}$ to each received sequence $y_2^n$.
The average probability of error is defined as,
\begin{align}\label{Pe1}
P_{e,1}^n &= \mbox{Pr}\{({\widehat W_{01}},{\widehat W_1},{\widehat W_2})\neq (W_0,W_1,W_2)\}\\\label{Pe2}
P_{e,2}^n &= \mbox{Pr}\{({\widehat W_{02}})\neq (W_0)\}\\
\label{Pe}
P_{e}^n &= \max\{P_{e,1}^n,P_{e,2}^n\}
\end{align}

The ignorance level of the eavesdropper (Receiver~2), with respect to the confidential messages $W_1$ and $W_2$, is measured by equivocation rate $(1/n)H({W_1},{W_2}|{Y_{2}^n})$.
\end{definition}
\begin{definition}\label{thirddefi} A rate tuple $({R_0},{R_1},{R_2})$ is said to be achievable for MAWC-CM, if for any $\varepsilon > 0$ there exists a $({2^{nR_0}},{2^{nR_1}},{2^{nR_2}},n)$ code which satisfies
\begin{align}
&P_e^n < \varepsilon \\
\label{TarifSecrecy}
&{R_1}+{R_2} - \frac{1}{n}H({W_1},{W_2}|{Y_{2}^n}) \leq \varepsilon
\end{align}
\end{definition}
for sufficiently large $n$. Note that secrecy requirement (\ref{TarifSecrecy}) implies:
\begin{equation}\label{Privatesecrecy}
{R_k} - \frac{1}{n}H({W_k}|{Y_{2}^n}) \leq \varepsilon \,\,\,\mbox{for}\,\,\,k=1,2
\end{equation}that also has been shown in \cite{cooperation}.
The secrecy capacity region of the MAWC-CM is defined as the closure of the set of all achievable rate tuples $(R_{0},R_{1},R_{2})$.
%\vspace{-2mm}
%%%%%%%%%%%%%%%%%%%%%%%%%%%%%%%%%%%%%%%%%%%%%%%%%%%%%%%%%%%%%%%%%%%%%%%%%%%%%%%%%%%%%%%%%%%%%%%%%%%%
%%%%%%%%%%%%%%%%%%%%%%%%%%%%%%%%%%%%%%%%%%%%%%%% MAIN RESULTS %%%%%%%%%%%%%%%%%%%%%%%%%%%%%%%%%%%%%%
\section{Discrete Memoryless MAWC-CM}\label{discreteinneroutersec}
In this section, we derive an outer bound on the secrecy capacity region of discrete memoryless MAWC-CM in Theorem~\ref{outerthm} and an inner bound in Theorem~\ref{achievthm}. We show that these bounds meet each other in a special case.
\subsection{Outer Bound}

\begin{theorem}\label{outerthm} (Outer bound) The secrecy capacity region of MAWC-CM is included in the set of rates satisfying
\begin{align}
    \label{R0}
    {R_0} &\le \min \{I(U;Y_{1}),I(U;Y_{2})\}  \\
    \label{R1}
    {R_1} &\le I({V_1};Y_{1}|U)-I({V_1};Y_{2}|U)  \\
    \label{R2}
    {R_2} &\le I({V_2};Y_{1}|U)-I({V_2};Y_{2}|U) \\
    \label{jameR1R2}
    {R_1+R_2} &\le I({V_1},{V_2};Y_{1}|U)-I({V_1},{V_2};Y_{2}|U)
%    \label{jameR0R1R2}
%    {R_0+R_1+R_2} &\le I({V_1},{V_2};Y_{1})-I({V_1},{V_2};Y_{2}|U)
\end{align}
for some joint distribution
\begin{equation}
p(u)p({v_1},{v_2}|u)p({x_1}|{v_1})p({x_2}|{v_2})p(y_{1},y_{2}|{x_1},{x_2})\label{DistOuter}
\end{equation}where the auxiliary random variables $U$, $V_1$ and $V_2$ are bounded in cardinality by
\begin{align}
  |{\cal U}| &\leq |{\cal X}_1|.|{\cal X}_2| + 7\\
  |{\cal V}_1| &\leq (|{\cal X}_1|.|{\cal X}_2| + 3).(|{\cal X}_1|.|{\cal X}_2| + 7)\\
  |{\cal V}_2| &\leq (|{\cal X}_1|.|{\cal X}_2| + 3).(|{\cal X}_1|.|{\cal X}_2| + 7).
\end{align}
\end{theorem}
\begin{proof}
See Appendix~\ref{outerproof}.
\end{proof}
\begin{remark}\label{firstrem} If transmitter~1 (or transmitter~2) does not send any messages, by setting $V_{1}=\emptyset$ (or $V_{2}=\emptyset$) in Theorem~\ref{outerthm} the region reduces to the capacity region of the broadcast channel with confidential messages discussed in \cite{CsiszarKorner}.
\end{remark}
%%%%%%%%%%%%%%%%%%%%%%%%%%%%%%%%%%%%%%%%%%%%%% INNER BOUND %%%%%%%%%%%%%%%%%%%%%%%%%%%%%%%%%%%%%%%
\subsection{Achievability}\label{achipersec}

\begin{theorem}\label{achievthm}(Achievability) For a discrete memoryless MAWC-CM, the secrecy rate region ${\cal R}({\pi _I})$ is achievable, where ${\cal R}({\pi _I})$ is the closure of the convex hull of all non-negative $({R_0},{R_1},{R_2})$ satisfying
\begin{align}\label{thm2asli}
%\left\{ \begin{array}{l}
{R_0} &\le I(U;Y_{2})\\\label{thm2asliR1}
{R_1} &\le I({V_1};Y_{1}|{V_2},U) - I({V_1};Y_{2}|U)\\\label{thm2asliR2}
{R_2} &\le I({V_2};Y_{1}|{V_1},U) - I({V_2};Y_{2}|U)\\\label{thm2asliR1R2}
{R_1} + {R_2} &\le I({V_1},{V_2};Y_{1}|U) - I({V_1},{V_2};Y_{2}|U)\,\\\label{thm2asliR0R1R2}
{R_0} + {R_1} + {R_2} &\le I({V_1},{V_2};Y_{1}) - I({V_1},{V_2};Y_{2}|U)
%\end{array} \right.
\end{align}and ${\pi _I}$ denotes the class of joint probability mass functions $p(u,{v_1},{v_2},{x_1},{x_2},y_{1},y_{2})$ that factor as
\begin{equation}\label{DistInner}
p(u)p(v_1|u)p(v_2|u)p({x_1}|v_1)p({x_2}|v_2)p(y_{1},y_{2}|{x_1},{x_2})
\end{equation}in which the auxiliary random variables $U$, $V_1$ and $V_2$ are bounded in cardinality by
\begin{align}
  |{\cal U}| &\leq |{\cal X}_1|.|{\cal X}_2| + 6\\
  |{\cal V}_1| &\leq (|{\cal X}_1|.|{\cal X}_2| + 5).(|{\cal X}_1|.|{\cal X}_2| + 6)\\
  |{\cal V}_2| &\leq (|{\cal X}_1|.|{\cal X}_2| + 5).(|{\cal X}_1|.|{\cal X}_2| + 6).
\end{align}
\end{theorem}
\begin{proof}\label{achievprof}
In the following, we provide outlines of our achievable scheme and the details of the proof are referred to Appendix~\ref{achievproof}. In the coding scheme we use superposition technique and Wyner's wiretap coding \cite{Wyner}, as the secrecy achievability method. We illustrate the common message $w_{0}$ with the auxiliary codeword $u^{n}$. All receivers are able to decode this codeword. Therefore, it does not need to be protected from the illegal user by Wyner's coding technique. The auxiliary codeword $v_{1}^{n}$, which illustrates the private message $w_{1}$, is superimposed on top of $u^{n}$ and is decoded only by receiver~1. Also, the auxiliary codeword $v_{2}^{n}$, which illustrates the message $w_{2}$, is superimposed on top of $u^{n}$ and is decoded only by receiver~1. These codewords are protected from the illegal user by Wyner's coding technique. The structure of the encoder is depicted in Fig.~\ref{divomin} in Appendix~\ref{achievproof}. Transmitted codewords $x_1^n$ and $x_2^n$ are drawn based on $v_1^n$ and $v_2^n$ respectively, according to (\ref{DistInner}).
\end{proof}
\begin{remark}
If we convert our model to a MAWC without common message, by setting $U= \emptyset$, $V_1=X_1$ and $V_2=X_2$ in Theorem~\ref{achievthm}, the region reduces to the achievable secrecy rate region of the MAWC without common message that is reported in \cite{cooperation} and its Gaussian version is firstly introduced in \cite{GMAWC} and \cite{GMAWCJamming}.
\end{remark}
\begin{remark}
If we convert the model to a broadcast channel with confidential messages, our region includes the region discussed by  Csisz\'{a}r and K\"{o}rner in \cite{CsiszarKorner}. It can be verified by setting $V_1=\emptyset$ or $V_2=\emptyset$ in (\ref{thm2asli})-(\ref{thm2asliR0R1R2}).
\end{remark}
\subsection{Switch Channel}
\begin{figure}%*}[ht]
\noindent
\makeatletter%
\if@twocolumn%
\centering
\includegraphics[width=8.87cm]{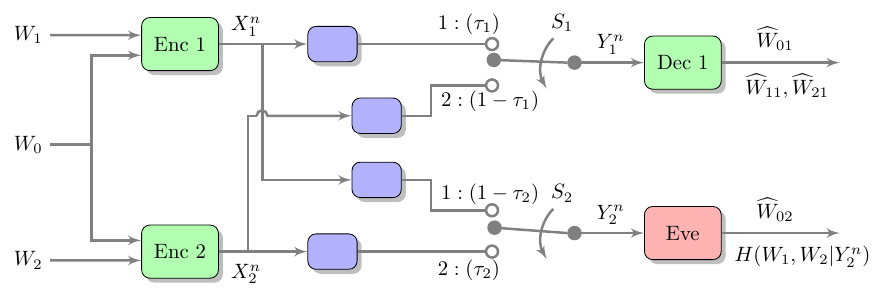}% asli 12.50cm and 8.87
\caption{The Switch channel model.}
\label{figIVI}
\else
\centering
\includegraphics[width=12.50cm]{SC2.pdf}% asli 12.50cm and 8.87
\caption{The Switch channel model.}
\label{figIVI}
\fi
\makeatother
\vspace{-0.8cm}
\end{figure}%*}
Now, we obtain the secrecy capacity region for a switch channel. In the switch channel (see Fig. \ref{figIVI}) the receivers cannot listen to both transmitters at the same time. For example, each receiver can listen to only one frequency whereas each transmitter can broadcast at various frequencies during the symbol time $i$. We assume that at each symbol time $i$, each receiver $t$ for $t\in\{1,2\}$ has access to a random switch $s_{t}\in\{1,2\}$, which independently is set to $t$ or $\bar{t}$ with probabilities
\begin{align}
P(S_{t,i}=t) &= \tau_{t}\\
P(S_{t,i}=\bar{t}) &= 1-\tau_{t},\,\,\,i=1,...,n
\end{align}where $\bar{t}$ is complement of $t$. Hence, if $S_{t,i}=t$, receiver $t$ for $t\in\{1,2\}$ listens to the signal sent by the transmitter~$t$ (i.e. $x_{t,i}$) and if $S_{t,i}=\bar{t}$, receiver~$t$ listens to the signal sent by the transmitter~$\bar{t}$ (i.e. $x_{\bar{t},i}$). The switch channel is investigated in \cite{liumaric} as a special case of Interference channel. We generalize the interpretation of a switch channel to our model as follows: Consider a MAC with a common message and an eavesdropper that the legal receiver (receiver~1) can listen to only one of the transmitters at each time instant that is determined by the first switch state. The illegal receiver (in terms of private messages) can eavesdrop only one of the transmitters which is determined by the second switch state. We also assume that both receivers have access to switch state information. Thus, we have
  \noindent
  \makeatletter%
  \if@twocolumn%
\begin{align}\label{SCmodel}
P&(y_{t,i}|x_{1,i},x_{2,i},s_{t,i})=P(y_{t,i}|x_{1,i})1(s_{t,i}=1)\nonumber\\
&+P(y_{t,i}|x_{2,i})1(s_{t,i}=2)=P(y_{t,i}|x_{s_{t,i},i})
\end{align}
\else
\begin{align}\label{SCmodel}
P&(y_{t,i}|x_{1,i},x_{2,i},s_{t,i})=P(y_{t,i}|x_{1,i})1(s_{t,i}=1) + P(y_{t,i}|x_{2,i})1(s_{t,i}=2)=P(y_{t,i}|x_{s_{t,i},i})
\end{align}
  \fi
  \makeatother
where $1(.)$ is the indicator function. The switch state information $\left\{ {{S_{t,i}}} \right\}_{i = 1}^n$ is an i.i.d. process known at receiver $t$. Therefore, we can assume that $s_{t,i}$ is a part of the channel output. In other words, we set
\begin{equation}\label{yt}
{y_{t,i}} \triangleq \{ {k_{t,i}},{s_{t,i}}\}
\end{equation}where $k_{t,i}$ indicates the received signal at receiver $t$. For the described switch channel, we have the following theorem for the secrecy capacity region.

%\begin{thm}\label{thsc}
\begin{theorem}\label{scthm}
For the switch channel with two confidential messages and one common message, the secrecy capacity region $\mathcal{C}_S$ is the union of all $(R_{0},R_{1},R_{2})$ satisfying
\begin{align}
\label{capacity R0}
{R_0} &\le I(U;Y_{2})\\
\label{capacity R1}
{R_1} &\le I({V_1};Y_{1}|U) - I({V_1};Y_{2}|U)\\
\label{capacity R2}
{R_2} &\le I({V_2};Y_{1}|U) - I({V_2};Y_{2}|U)\\
\label{capacity R1+R2}
{R_1} + {R_2} &\le I({V_1},{V_2};Y_{1}|U) - I({V_1},{V_2};Y_{2}|U)\\
\label{capacity R0+R1+R2}
{R_0} + {R_1} + {R_2} &\le I({V_1},{V_2};Y_{1}) - I({V_1},{V_2};Y_{2}|U)%\nonumber
\end{align}over all distributions
\begin{equation}
p(u)p(v_1|u)p(v_2|u)p({x_1}|v_1)p({x_2}|v_2)p(y_{1},y_{2}|{x_1},{x_2})
\end{equation}where the auxiliary random variables $U$, $V_1$ and $V_2$ are bounded in cardinality by
\begin{align}
  |{\cal U}| &\leq |{\cal X}_1|.|{\cal X}_2| + 6\\
  |{\cal V}_1| &\leq (|{\cal X}_1|.|{\cal X}_2| + 4).(|{\cal X}_1|.|{\cal X}_2| + 6)\\
  |{\cal V}_2| &\leq (|{\cal X}_1|.|{\cal X}_2| + 4).(|{\cal X}_1|.|{\cal X}_2| + 6).
\end{align}
\end{theorem}

\begin{proof}
See Appendix~\ref{SCproof}.
\end{proof}

%\vspace{-3mm}
\section{Gaussian MAWC-CM}\label{gaussiansec}
\begin{figure}%*}[ht]
  \noindent
  \makeatletter%
  \if@twocolumn%
  \centering
  \includegraphics[width=8.50cm]{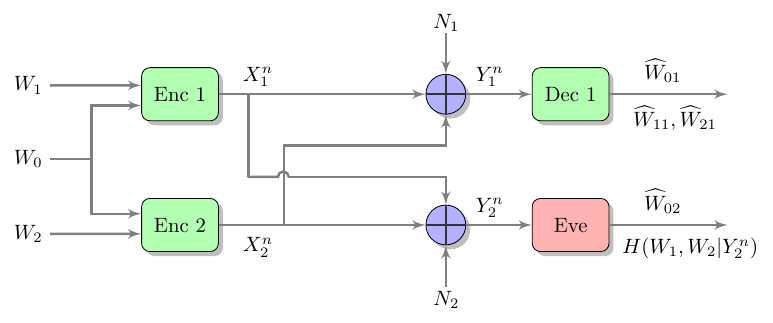}% asli 12.50cm 8.50
  \caption{The Gaussian MAWC-CM.}
  \label{figdiggaussi}
  \else
  \centering
  \includegraphics[width=12.50cm]{GaussianDiagram.pdf}% asli 12.50cm 8.50
  \caption{The Gaussian MAWC-CM.}
  \label{figdiggaussi}
  \fi
  \makeatother
\end{figure}%*}
\setlength{\textfloatsep}{10pt plus 1.0pt minus 2.0pt}
In this section, we consider Gaussian MAWC-CM as shown in Fig. \ref{figdiggaussi}, and derive inner and outer bounds on its secrecy capacity region. Relationships between the inputs and outputs of the channel, as shown in Fig.~\ref{figdiggaussi}, are given by

\begin{align}
\label{Y1}
Y_{1} &= X_{1} + X_{2} + N_{1}\\
\label{Y2}
Y_{2} &= X_{1} + X_{2} + N_{2}
%\label{V1G}
%X_{1} &=V_{1}\\
%\label{V2G}
%X_{2} &= V_{2}
\end{align}where $N_{1}$ and $N_{2}$ are independent zero-mean Gaussian Random Variables (RVs), with variances $\sigma_1^2$ and $\sigma_2^2$, and independent of the RVs $X_1$, $X_2$. We impose the power constraints $\frac{1}{n}\sum\nolimits_{i = 1}^n {E[X_{j,i}^2]} \le {P_j},\,\,\,j = 1,2$.
\subsection{Outer Bound}
\begin{theorem}\label{outergausithm} (Outer bound) The secrecy capacity region for the Gaussian MAWC-CM is included in the set of rates satisfying
  \noindent
  \makeatletter%
  \if@twocolumn%
\begin{equation}\label{outer}
\bigcup \left\{ \begin{array}{l}
{R_0} \le \min \{ %\begin{array}{l}
C (\frac{{(1 - {\beta _1}){P_1} + (1 - {\beta _2}){P_2} + (1 - {\beta _1}{\beta _2}){\rho \sqrt {{P_1}{P_2}}}}}{{{\beta _1}{P_1} + {\beta _2}{P_2} + 2{\beta _1}{\beta _2}{\rho \sqrt {{P_1}{P_2}}} + \sigma _1^2}})\\
\,\,\,\,\,,C (\frac{{(1 - {\beta _1}){P_1} + (1 - {\beta _2}){P_2} + (1 - {\beta _1}{\beta _2}){\rho \sqrt {{P_1}{P_2}}}}}{{{\beta _1}{P_1} + {\beta _2}{P_2} + 2{\beta _1}{\beta _2}{\rho \sqrt {{P_1}{P_2}}} + \sigma _2^2}})\}\\
%\end{array} \}\\
%{R_1} \le \frac{1}{2}\log \left( {1 + \frac{{{\beta _1}{Q_1} + {\beta _2}{Q_2} + 2{\beta _1}{\beta _2}{Q_3}}}{{\sigma _1^2}}} \right) \nonumber\\
%\,\,\,\,\,\,\,\,\,\,- \frac{1}{2}\log \left( {1 + \frac{{{\beta _1}Q_1^2 + 2{\beta _1}{\beta _2}{Q_1}{Q_3} + Q_3^2 - (1 - {\beta _2}){Q_1}{Q_2}}}{{{Q_1}{Q_2} - Q_3^2 + \sigma _2^2{Q_1}}}} \right)\\
%{R_2} \le {\left[ {\frac{1}{2}\log \left( {1 + \frac{{{\beta _1}{Q_1} + {\beta _2}{Q_2} + 2{\beta _1}{\beta _2}{Q_{_3}}}}{{\sigma _1^2}}} \right) - \frac{1}{2}\log \left( {1 + \frac{{{\beta _2}Q_2^2 + 2{\beta _1}{\beta _2}{Q_2}{Q_3} + Q_3^2 - (1 - {\beta _1}){Q_1}{Q_2}}}{{{Q_1}{Q_2} - Q_3^2 + \sigma _2^2{Q_2}}}} \right)} \right]^ + }\\
{R_1} + {R_2} \le {[C \left( {\frac{{{\beta _1}{P_1} + {\beta _2}{P_2} + 2{\beta _1}{\beta _2}{\rho \sqrt {{P_1}{P_2}}}}}{{\sigma _1^2}}} \right) }\\
\,\,\,\,- C \left( {\frac{{{\beta _1}{P_1} + {\beta _2}{P_2} + 2{\beta _1}{\beta _2}{\rho \sqrt {{P_1}{P_2}}}}}{{\sigma _2^2}}} \right)]^ + \\
%{R_0} + {R_1} + {R_2} \le {[C \left( {\frac{{{P_1} + {P_2} + 2{\rho \sqrt {{P_1}{P_2}}}}}{{\sigma _1^2}}} \right) }\\
%\,\,\,\,- C \left( {\frac{{{\beta _1}{P_1} + {\beta _2}{P_2} + 2{\beta _1}{\beta _2}{\rho \sqrt {{P_1}{P_2}}}}}{{\sigma _2^2}}} \right)]^ + \\
\end{array} \right.
\end{equation}
\else
\begin{equation}\label{outer}
\bigcup \left\{ \begin{array}{l}
{R_0} \le \min \{ %\begin{array}{l}
C (\frac{{(1 - {\beta _1}){P_1} + (1 - {\beta _2}){P_2} + (1 - {\beta _1}{\beta _2}){\rho \sqrt {{P_1}{P_2}}}}}{{{\beta _1}{P_1} + {\beta _2}{P_2} + 2{\beta _1}{\beta _2}{\rho \sqrt {{P_1}{P_2}}} + \sigma _1^2}}),C (\frac{{(1 - {\beta _1}){P_1} + (1 - {\beta _2}){P_2} + (1 - {\beta _1}{\beta _2}){\rho \sqrt {{P_1}{P_2}}}}}{{{\beta _1}{P_1} + {\beta _2}{P_2} + 2{\beta _1}{\beta _2}{\rho \sqrt {{P_1}{P_2}}} + \sigma _2^2}})\}\\
%\end{array} \}\\
%{R_1} \le \frac{1}{2}\log \left( {1 + \frac{{{\beta _1}{Q_1} + {\beta _2}{Q_2} + 2{\beta _1}{\beta _2}{Q_3}}}{{\sigma _1^2}}} \right) \nonumber\\
%\,\,\,\,\,\,\,\,\,\,- \frac{1}{2}\log \left( {1 + \frac{{{\beta _1}Q_1^2 + 2{\beta _1}{\beta _2}{Q_1}{Q_3} + Q_3^2 - (1 - {\beta _2}){Q_1}{Q_2}}}{{{Q_1}{Q_2} - Q_3^2 + \sigma _2^2{Q_1}}}} \right)\\
%{R_2} \le {\left[ {\frac{1}{2}\log \left( {1 + \frac{{{\beta _1}{Q_1} + {\beta _2}{Q_2} + 2{\beta _1}{\beta _2}{Q_{_3}}}}{{\sigma _1^2}}} \right) - \frac{1}{2}\log \left( {1 + \frac{{{\beta _2}Q_2^2 + 2{\beta _1}{\beta _2}{Q_2}{Q_3} + Q_3^2 - (1 - {\beta _1}){Q_1}{Q_2}}}{{{Q_1}{Q_2} - Q_3^2 + \sigma _2^2{Q_2}}}} \right)} \right]^ + }\\
{R_1} + {R_2} \le {[C \left( {\frac{{{\beta _1}{P_1} + {\beta _2}{P_2} + 2{\beta _1}{\beta _2}{\rho \sqrt {{P_1}{P_2}}}}}{{\sigma _1^2}}} \right) }- C \left( {\frac{{{\beta _1}{P_1} + {\beta _2}{P_2} + 2{\beta _1}{\beta _2}{\rho \sqrt {{P_1}{P_2}}}}}{{\sigma _2^2}}} \right)]^+
%\\
%{R_0} + {R_1} + {R_2} \le {[C \left( {\frac{{{P_1} + {P_2} + 2{\rho \sqrt {{P_1}{P_2}}}}}{{\sigma _1^2}}} \right) }- C \left( {\frac{{{\beta _1}{P_1} + {\beta _2}{P_2} + 2{\beta _1}{\beta _2}{\rho \sqrt {{P_1}{P_2}}}}}{{\sigma _2^2}}} \right)]^ + \\
\end{array} \right.
\end{equation}
\fi
\makeatother
where $C(x) = (1/2)\log (1 + x)$ and the union is taken over all $0 \le \beta _1 \le {1}$, $0 \le \beta _2 \le {1}$ and $0 \le \rho \le {1}$.
\end{theorem}
\begin{proof}
See Appendix~\ref{outergausiproof}.
\end{proof}
\subsection{Inner Bound}
\begin{theorem}\label{achievgausithm} (Achievability) An inner bound on the secrecy capacity region of Gaussian MAWC-CM is:
\noindent
\makeatletter%
\if@twocolumn%
\begin{equation}\label{GAchievable}
\bigcup \left\{ \begin{array}{l}
{R_0} \le C(\frac{{\beta _1^2{P_1} + \beta _2^2{P_2} + 2{\beta _1}{\beta _2}\sqrt {{P_1}{P_2}} }}{{(1 - \beta _1^2){P_1} + (1 - \beta _2^2){P_2} + \sigma _2^2}})\\
{R_1} \le C(\frac{{(1 - \beta _1^2){P_1}}}{{\sigma _1^2}}) - C(\frac{{(1 - \beta _1^2){P_1}}}{(1 - \beta _2^2){P_2}+{\sigma _2^2}})\\
{R_2} \le C(\frac{{(1 - \beta _2^2){P_2}}}{{\sigma _1^2}}) - C(\frac{{(1 - \beta _2^2){P_2}}}{(1 - \beta _1^2){P_1}+{\sigma _2^2}})\\
{R_1} + {R_2} \le C(\frac{{(1 - \beta _1^2){P_1} + (1 - \beta _2^2){P_2}}}{{\sigma _1^2}}) \\
\,\,\,\,- C(\frac{{(1 - \beta _1^2){P_1} + (1 - \beta _2^2){P_2}}}{{\sigma _2^2}})\\
{R_0} + {R_1} + {R_2} \le C(\frac{{{P_1} + {P_2} + 2{\beta _1}{\beta _2}\sqrt {{P_1}{P_2}} }}{{\sigma _1^2}}) \\
\,\,\,\,- C(\frac{{(1 - \beta _1^2){P_1} + (1 - \beta _2^2){P_2}}}{{\sigma _2^2}})
\end{array} \right.
\end{equation}
\else
\begin{equation}\label{GAchievable}
\bigcup \left\{ \begin{array}{l}
{R_0} \le C(\frac{{\beta _1^2{P_1} + \beta _2^2{P_2} + 2{\beta _1}{\beta _2}\sqrt {{P_1}{P_2}} }}{{(1 - \beta _1^2){P_1} + (1 - \beta _2^2){P_2} + \sigma _2^2}})\\
{R_1} \le C(\frac{{(1 - \beta _1^2){P_1}}}{{\sigma _1^2}}) - C(\frac{{(1 - \beta _1^2){P_1}}}{(1 - \beta _2^2){P_2}+{\sigma _2^2}})\\
{R_2} \le C(\frac{{(1 - \beta _2^2){P_2}}}{{\sigma _1^2}}) - C(\frac{{(1 - \beta _2^2){P_2}}}{(1 - \beta _1^2){P_1}+{\sigma _2^2}})\\
{R_1} + {R_2} \le C(\frac{{(1 - \beta _1^2){P_1} + (1 - \beta _2^2){P_2}}}{{\sigma _1^2}}) - C(\frac{{(1 - \beta _1^2){P_1} + (1 - \beta _2^2){P_2}}}{{\sigma _2^2}})\\
{R_0} + {R_1} + {R_2} \le C(\frac{{{P_1} + {P_2} + 2{\beta _1}{\beta _2}\sqrt {{P_1}{P_2}} }}{{\sigma _1^2}}) - C(\frac{{(1 - \beta _1^2){P_1} + (1 - \beta _2^2){P_2}}}{{\sigma _2^2}})
\end{array} \right.
\end{equation}
\fi
\makeatother
where $C(x) = (1/2)\log (1 + x)$ and the union is taken over all $0 \le \beta _1 \le {1}$ and $0 \le \beta _2 \le {1}$.
\end{theorem}
\begin{proof}\label{achievgausithmprof}
The achievable rate region in Theorem~\ref{achievthm} can be extended to the discrete-time Gaussian memoryless case with continuous alphabets by standard arguments \cite{CoverThomas}. Hence, it is sufficient to evaluate (\ref{thm2asli})-(\ref{thm2asliR0R1R2}) with appropriate choice of input distribution to reach (\ref{GAchievable}). We constrain all the inputs to be Gaussian. For certain $0\leq \beta_{1}\leq 1$ and $0\leq \beta_{2}\leq 1$ consider the following mapping in (\ref{V1})-(\ref{X2}) for the generated codebook in Theorem~\ref{achievthm} with respect to the p.m.f (\ref{DistInner}), which contains the Gaussian version of the superposition coding and random binning:
\begin{align}
\label{V1}
{V_1} &= \sqrt {{P_1}} {\beta _1}U + \sqrt {{P_1}(1 - {\beta_1^2})} {K_1}\\
\label{X1}
X_1 &= V_1\\
\label{V2}
{V_2} &= \sqrt {{P_2}} {\beta _2}U + \sqrt {{P_2}(1 - {\beta_2^2})} {K_2}\\
\label{X2}
X_2 &= V_2.
\end{align}where $U$, $K_1$ and $K_2$ are independent, zero-mean and unit variance Gaussian RVs. Using the above mapping with the channel model in (\ref{Y1})-(\ref{Y2}), and by calculating mutual information functions for Gaussian RVs similar to the method in \cite{CoverThomas}, we have:
\begin{align}
I(U;Y_{2})&=C(\frac{{\beta _1^2{P_1} + \beta _2^2{P_2} + 2{\beta _1}{\beta _2}\sqrt {{P_1}{P_2}} }}{{(1 - \beta _1^2){P_1} + (1 - \beta _2^2){P_2} + \sigma _2^2}})\label{IUY2}\\
I({V_1};Y_{1}|{V_2},U)&= C(\frac{{(1 - \beta _1^2){P_1}}}{{\sigma _1^2}})\label{IV1Y11V2U}\\
I({V_1};Y_{2}|U)&= C(\frac{{(1 - \beta _1^2){P_1}}}{(1 - \beta _2^2){P_2}+{\sigma _2^2}})\label{IV1Y21U}\\
I({V_2};Y_{1}|{V_1},U)&= C(\frac{{(1 - \beta _2^2){P_2}}}{{\sigma _1^2}})\label{IV2Y11V1U}\\
I({V_2};Y_{2}|U)&=C(\frac{{(1 - \beta _2^2){P_2}}}{(1 - \beta _1^2){P_1}+{\sigma _2^2}})\label{IV2Y21U}\\
I({V_1},{V_2};Y_{1}|U)&= C(\frac{{(1 - \beta _1^2){P_1} + (1 - \beta _2^2){P_2}}}{{\sigma _1^2}})\label{IV1V2Y11U}\\
I({V_1},{V_2};Y_{2}|U)&= C(\frac{{(1 - \beta _1^2){P_1} + (1 - \beta _2^2){P_2}}}{{\sigma _2^2}})\label{IV1V2Y21U}\\
I({V_1},{V_2};Y_{1})&=C(\frac{{{P_1} + {P_2} + 2{\beta _1}{\beta _2}\sqrt {{P_1}{P_2}} }}{{\sigma _1^2}})\label{IV1V2Y1}
\end{align}where $C(x) = (1/2)\log (1 + x)$. Considering Theorem 2 and (\ref{IUY2})-(\ref{IV1V2Y1}) completes the proof.
\end{proof}
As mentioned in the Introduction section we aim to compare our derived bounds with each other and also with the capacity region of the Gaussian compound MAC. Hence, we first derive this region as follows.

\begin{theorem}\label{cmacthm}
The capacity region of Gaussian compound MAC with common information is given by:
%%%%%%%%%%%%%%%%%%%%%%%%%%%%%%%%my array%%%%%%%%%%%%%%%%
%%%%%%%%%%%%%%%%%%%%%%%%%%%%%%%%%%%%%%%%%%%%%%%%%%%%%%%%
%\[\bigcup {\left\{ \begin{array}{l}
\noindent
\makeatletter%
\if@twocolumn%
\begin{equation}\label{capacityCMAC}%%%ezafe shod
\bigcup {\left\{ \begin{array}{l}
{R_1} \le \min \{ C(\frac{{{P_1}(1 - \beta _1^2)}}{{\sigma _1^2}}),C(\frac{{{P_1}(1 - \beta _1^2)}}{{\sigma _2^2}})\} \\
{R_2} \le \min \{ C(\frac{{{P_2}(1 - \beta _2^2)}}{{\sigma _1^2}}),C(\frac{{{P_2}(1 - \beta _2^2)}}{{\sigma _2^2}})\} \\
{R_1} + {R_2} \le \min \{ C(\frac{{{P_1}(1 - \beta _1^2) + {P_2}(1 - \beta _2^2)}}{{\sigma _1^2}})\\
\,\,\,\,,C(\frac{{{P_1}(1 - \beta _1^2) + {P_2}(1 - \beta _2^2)}}{{\sigma _2^2}})\} \\
{R_0} + {R_1} + {R_2} \le \min \{ C(\frac{{{P_1} + {P_2} + 2\sqrt {{P_1}{P_2}} {\beta _1}{\beta _2}}}{{\sigma _1^2}})\\
\,\,\,\,,C(\frac{{{P_1} + {P_2} + 2\sqrt {{P_1}{P_2}} {\beta _1}{\beta _2}}}{{\sigma _2^2}})\}
\end{array} \right.}
\end{equation}
\else
\begin{equation}\label{capacityCMAC}%%%ezafe shod
\bigcup {\left\{ \begin{array}{l}
{R_1} \le \min \{ C(\frac{{{P_1}(1 - \beta _1^2)}}{{\sigma _1^2}}),C(\frac{{{P_1}(1 - \beta _1^2)}}{{\sigma _2^2}})\} \\
{R_2} \le \min \{ C(\frac{{{P_2}(1 - \beta _2^2)}}{{\sigma _1^2}}),C(\frac{{{P_2}(1 - \beta _2^2)}}{{\sigma _2^2}})\} \\
{R_1} + {R_2} \le \min \{ C(\frac{{{P_1}(1 - \beta _1^2) + {P_2}(1 - \beta _2^2)}}{{\sigma _1^2}})
,C(\frac{{{P_1}(1 - \beta _1^2) + {P_2}(1 - \beta _2^2)}}{{\sigma _2^2}})\} \\
{R_0} + {R_1} + {R_2} \le \min \{ C(\frac{{{P_1} + {P_2} + 2\sqrt {{P_1}{P_2}} {\beta _1}{\beta _2}}}{{\sigma _1^2}})
,C(\frac{{{P_1} + {P_2} + 2\sqrt {{P_1}{P_2}} {\beta _1}{\beta _2}}}{{\sigma _2^2}})\}
\end{array} \right.}
\end{equation}
\fi
\makeatother
where $C(x) = (1/2)\log (1 + x)$ and the union is taken over all $0 \le \beta _1 \le {1}$ and $0 \le \beta _2 \le {1}$.$\\$
\end{theorem}
\begin{proof}\label{achievproofthm}
It is clear that to obtain this capacity region, Propositions~6.1 and 6.2 in \cite{simeone}, which are outer and inner bounds on the capacity region of the Gaussian compound MAC with conferencing links, can be modified by setting $C_{12}=C_{21}=0$ in them (i.e., ignoring conferencing links) and by adopting them to our defined channel parameters in (\ref{Y1}) and (\ref{Y2}).
\end{proof}
%\vspace{-2mm}
%%%%%%%%%%%%%%%%%%%%%%%%%%%%%%%%%%%%%%%%%%%%%%% eafe shode az ICC %%%%%%%%%%%%%%%%%%%
%%%%%%%%%%%%%%%%%%%%%%%%%%%%%%%%%%%%%%%%%%%%%%%%%%%%%%%%%%%%%%%%%%%%%%%%%%%%%%%%%%%%
\subsection{Examples}
%\vspace{-1.5mm}
In this part, we provide numerical examples and compare our derived inner and outer bounds on the secrecy capacity region of Gaussian MAWC-CM. We also compare these bounds with the capacity region of the Gaussian compound MAC illustrated in (\ref{capacityCMAC}).
As an example, for the values $P_{1}=P_{2}=1$, $\sigma_1^2=0.1$ and $\sigma_2^2=0.3$ the outer bound in Theorem~\ref{outergausithm} and the achievable rate region in Theorem~\ref{achievgausithm} are depicted in Fig.~\ref{figII}. In order to illustrate the effect of secrecy constraint and noise power on the rate region of MAWC-CM, we also compare our derived regions with the capacity region of the Gaussian compound MAC in (\ref{capacityCMAC}). These comparisons are shown in Figures \ref{figIII} and \ref{figIV}. Actually, in the Compound MAC (CMAC) both receivers should decode $W_{0},W_{1},W_{2}$ reliably, while in the defined MAWC-CM model the messages $W_1$ and $W_2$ should be kept secret from receiver~2. As it can be seen in Fig.~\ref{figIII} for channel parameters $P_{1}=P_{2}=1$, $\sigma_1^2=0.1$ and $\sigma_2^2=0.3$ (i.e., the same as for Fig.~\ref{figII}) the achievable rates and outer bounds on $R_1$ and $R_2$ (rate of $W_1$ and $W_2$ respectively, which are decoded by receiver~1) for MAWC-CM is less than that for CMAC due to secrecy constraint for decoding messages $W_1$ and $W_2$. Note that in Figures \ref{figIII} and \ref{figIV} we present the regions in two-dimensional by projecting on $R_0$ plane to have a better illustration. Based on (\ref{capacityCMAC}) it is clear that if the noise power of receiver~2 (i.e., $\sigma_2^2$) increases, the capacity region of Gaussian CMAC may remain as before or decreases (i.e., it  does not increase). On the other hand, there exist scenarios for MAWC-CM (see (\ref{GAchievable})) for which increasing $\sigma_2^2$ results in increasing its achievable rate region. For comparison, assume changing $\sigma _2^2=0.3$ to $\sigma _2^2=0.6$ in the above example. As it can be seen in Fig. \ref{figIV}, the achievable rate region of MAWC-CM is larger than the capacity region of CMAC for the new parameters. This can be interpreted as follows: the transmitted signals from transmitters~1 and 2 are extremely attenuated at the receiver~2 in comparison to the investigated case shown in Fig. \ref{figIII}. So, for this case the requirement of secrecy of $W_1$ and $W_2$ from receiver~2  in MAWC-CM can increase the achievable rate region in comparison with that of CMAC wherein $W_1$ and $W_2$ should be reliably decoded  by receiver~2.
\begin{figure}%*}[ht]
  \noindent
  \makeatletter%
  \if@twocolumn%
  \centering
  \includegraphics[width=8.50cm]{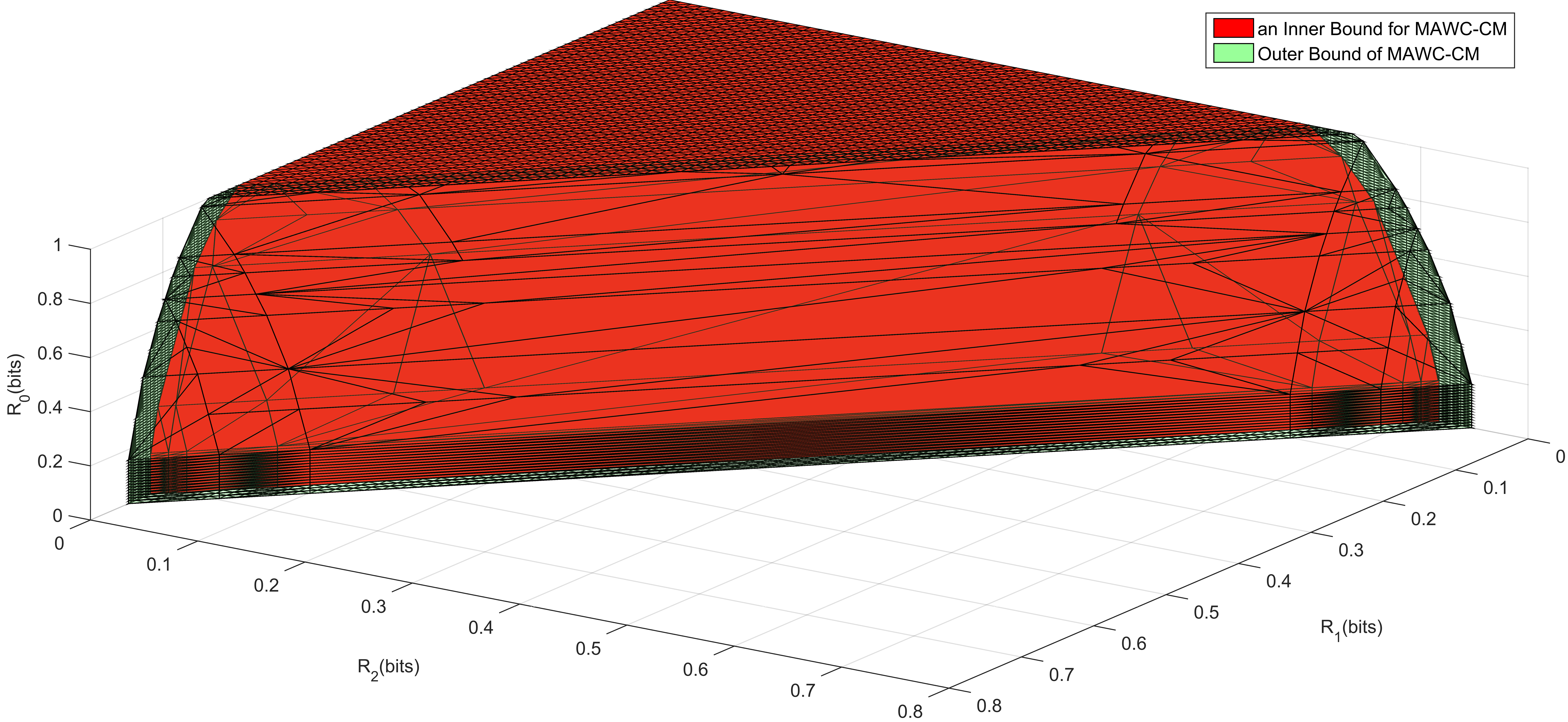}% asli 12.50cm 8.50
  \caption{\small{Achievable rate region and Outer bound of Gaussian MAWC-CM for $P_1=P_2=1$, $\sigma _1^2=0.1$ and $\sigma _2^2=0.3$.}}
  \label{figII}
  \else
  \centering
  \includegraphics[width=12.50cm]{3InnerandOuter321.png}% asli 12.50cm 8.50
  \caption{\small{Achievable rate region and Outer bound of Gaussian MAWC-CM for $P_1=P_2=1$, $\sigma _1^2=0.1$ and $\sigma _2^2=0.3$.}}
  \label{figII}
  \fi
  \makeatother
\end{figure}%*}
%%%%%%%%%%% End of Omitting figII
%\setlength{\textfloatsep}{10pt plus 1.0pt minus 2.0pt}
\begin{figure}%*}[ht]
  \noindent
  \makeatletter%
  \if@twocolumn%
  \centering
  \includegraphics[width=8.50cm]{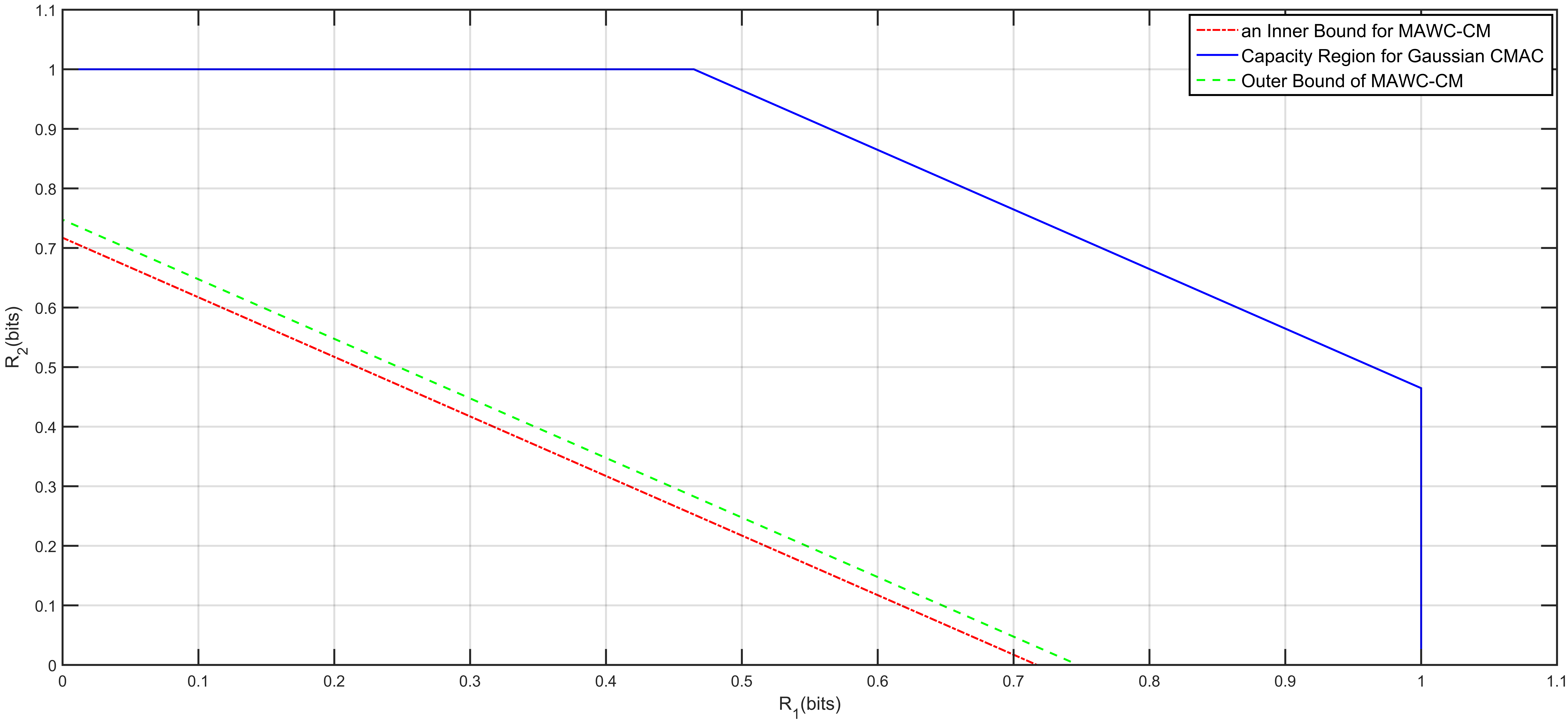}% asli 12.50cm 8.50
  \caption{\small{Achievable rate region and Outer bound of Gaussian MAWC-CM and the Capacity region of Gaussian CMAC for $P_1=P_2=1$, $\sigma _1^2=0.1$ and $\sigma _2^2=0.3$.}}
  \label{figIII}
  \else
  \centering
  \includegraphics[width=12.50cm]{2InnerOuterCMAC2.png}% asli 12.50cm 8.50
  \caption{\small{Achievable rate region and Outer bound of Gaussian MAWC-CM and the Capacity region of Gaussian CMAC for $P_1=P_2=1$, $\sigma _1^2=0.1$ and $\sigma _2^2=0.3$.}}
  \label{figIII}
  \fi
  \makeatother
\end{figure}%*}
\begin{figure}[!ht]%*}[ht]
  \noindent
  \makeatletter%
  \if@twocolumn%
  \centering
  \includegraphics[width=8.50cm]{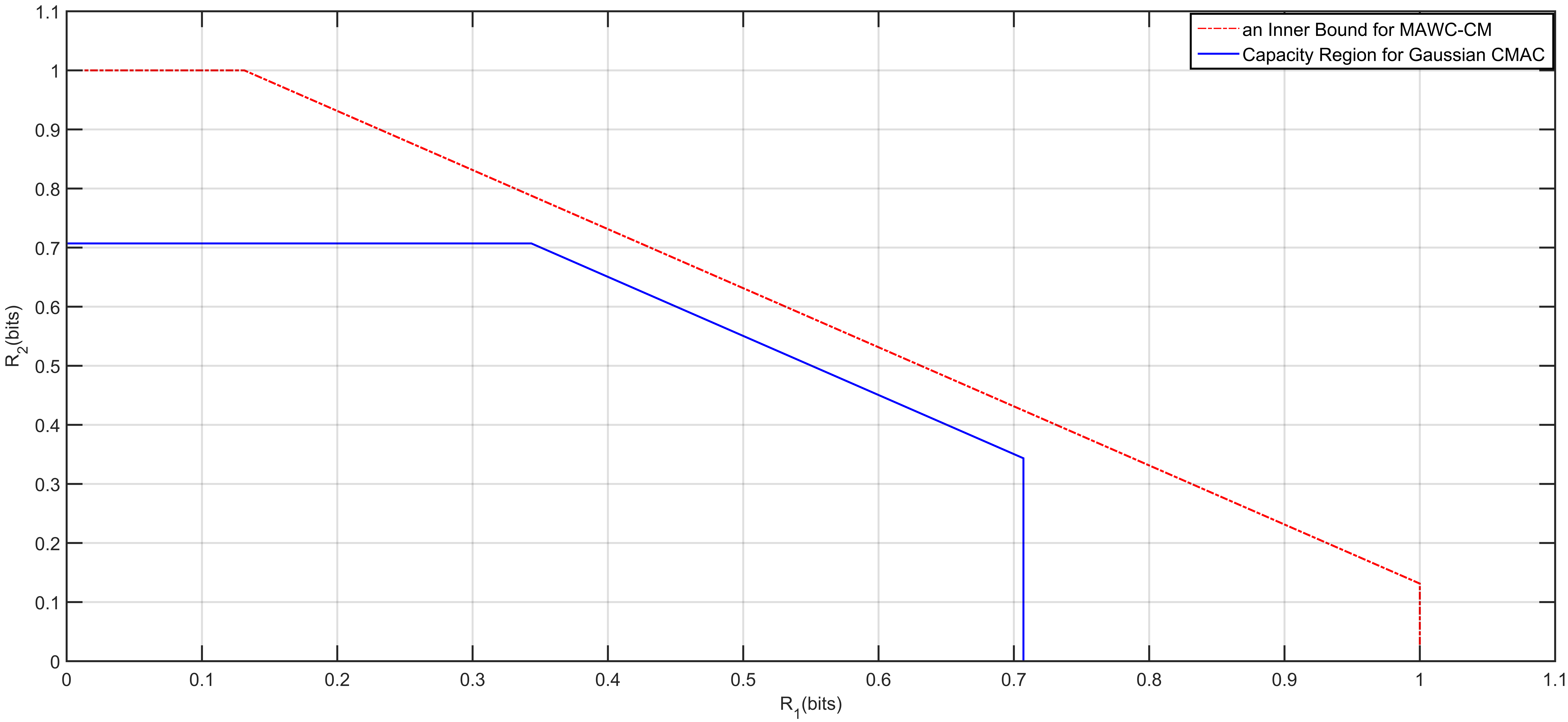}% asli 12.50cm 8.50
  \caption{\small{Achievable rate region of Gaussian MAWC-CM and the Capacity region of Gaussian CMAC for $P_1=P_2=1$, $\sigma _1^2=0.1$ and $\sigma _2^2=0.6$.}}
  \label{figIV}
  \else
  \centering
  \includegraphics[width=12.50cm]{2InnerCMAC32.png}% asli 12.50cm 8.50
  \caption{\small{Achievable rate region of Gaussian MAWC-CM and the Capacity region of Gaussian CMAC for $P_1=P_2=1$, $\sigma _1^2=0.1$ and $\sigma _2^2=0.6$.}}
  \label{figIV}
  \fi
  \makeatother
\end{figure}%*}
\section{Conclusions}\label{conclusions}
%\vspace{-1.5mm}
In this paper, we have studied the secrecy capacity region of Multiple Access Wire-tap Channel with Common Message (MAWC-CM). We have obtained inner and outer bounds on the secrecy capacity region for the general MAWC-CM and showed that these bounds meet each other for the switch channel model. As well, we have studied Gaussian MAWC-CM and derived inner and outer bounds on its secrecy capacity region. Providing numerical examples for the Gaussian case, we have illustrated the impact of noise and secrecy constraint on the capacity region. We have shown that there are scenarios for which the secret transmissions may increase achievable rate region in compare with the case that requires receiver~2 to decode the private messages.
%keeping some of the transmitted messages secret from the illegal user can increase the achievable rate region.
\section{Acknowledgment}
This work was partially supported by Iran NSF under Grant number 92-32575 and by Iran Telecom Research Center (ITRC).
\appendices
\section{Proof for Theorem~\ref{outerthm}}\label{outerproof}
The basis for deriving our outer bound is using Fano's inequality \cite{ElgamalKim} and applying the techniques in \cite{CsiszarKorner} to consider the secrecy constraints.

%We next show that any achievable rate tuples satisfy (\ref{R0})-(\ref{jameR0R1R2}) for some distribution factorized as (\ref{DistOuter}).
Consider a $(2^{nR_{0}},2^{nR_{1}},2^{nR_{2}},n,P_e^n)$ code for the MAWC-CM.
Applying Fano's inequality results in
\begin{align}
   &H({W_0},{W_1},{W_2}|{Y_{1}^n}) \le n{\varepsilon _1}\label{fano1}\\\label{fano2}
   &H({W_0}|{Y_{2}^n}) \le n{\varepsilon _2}.
\end{align}
where $\varepsilon_i\rightarrow 0,\,\,i=1,2$ as $P_e^n\rightarrow 0$.

In order to derive the upper bound on $R_1$, we first present the bound on $H({W_1}|{Y_{2}^n})$ as follows:
\noindent
\makeatletter%
\if@twocolumn%
\begin{align}
&H({W_1}|Y_{2}^n) = H({W_1}|Y_{2}^n,{W_0}) + I({W_1};{W_0}|Y_{2}^n)\nonumber\\
& = H({W_1}|Y_{2}^n,{W_0}) + H({W_0}|Y_{2}^n)-H({W_0}|Y_{2}^n,{W_1})\nonumber\\
& \mathop  \le \limits^{(a)} H({W_1}|Y_{2}^n,{W_0}) + n{\varepsilon _2}\nonumber\\
& \mathop  \le \limits^{(b)} H({W_1}|Y_{2}^n,{W_0}) - H({W_1}|{Y_{1}^n},{W_0}) + n({\varepsilon _1} + {\varepsilon _2})\nonumber\\
%\end{align}
%\begin{align}
& = I({W_1};{Y_{1}^n}|{W_0}) - I({W_1};Y_{2}^n|{W_0}) + n({\varepsilon _1} + {\varepsilon _2})\nonumber\\
%\end{align}
%\begin{align}
& = \sum\limits_{i = 1}^n {[I({W_1};{Y_{1,i}}|{W_0},{Y_{1}^{i - 1}})} - I({W_1};{Y_{2,i}}|{W_0},Y_{2,i + 1}^n)] \nonumber\\
&\,\,\,\,+ n({\varepsilon _1} + {\varepsilon _2}) \nonumber\\
&\mathop = \limits^{(c)} \sum \limits_{i = 1}^n {[I({W_1};{Y_{1,i}}|{W_0},{Y_{1}^{i - 1}},Y_{2,i + 1}^n) } \nonumber\\
&\,\,\,\, - I({W_1};{Y_{2,i}}|{W_0},{Y_{1}^{i - 1}},Y_{2,i + 1}^n)]  + n\varepsilon^\prime\nonumber\\\label{ghabli1}
&\mathop  = \limits^{(d)} \sum\limits_{i = 1}^n {[I({V_{1,i}};{Y_{1,i}}|{U_i}) - I({V_{1,i}};{Y_{2,i}}|{U_i})]}  + n\varepsilon^\prime
\end{align}
\else
\begin{align}
H({W_1}|Y_{2}^n) &= H({W_1}|Y_{2}^n,{W_0}) + I({W_1};{W_0}|Y_{2}^n)\nonumber\\
& = H({W_1}|Y_{2}^n,{W_0}) + H({W_0}|Y_{2}^n)-H({W_0}|Y_{2}^n,{W_1})\nonumber\\
& \mathop  \le \limits^{(a)} H({W_1}|Y_{2}^n,{W_0}) + n{\varepsilon _2}\nonumber\\
& \mathop  \le \limits^{(b)} H({W_1}|Y_{2}^n,{W_0}) - H({W_1}|{Y_{1}^n},{W_0}) + n({\varepsilon _1} + {\varepsilon _2})\nonumber\\
%\end{align}
%\begin{align}
& = I({W_1};{Y_{1}^n}|{W_0}) - I({W_1};Y_{2}^n|{W_0}) + n({\varepsilon _1} + {\varepsilon _2})\nonumber\\
& = \sum\limits_{i = 1}^n {[I({W_1};{Y_{1,i}}|{W_0},{Y_{1}^{i - 1}})} - I({W_1};{Y_{2,i}}|{W_0},Y_{2,i + 1}^n)] + n({\varepsilon _1} + {\varepsilon _2}) \nonumber\\
&\mathop = \limits^{(c)} \sum \limits_{i = 1}^n {[I({W_1};{Y_{1,i}}|{W_0},{Y_{1}^{i - 1}},Y_{2,i + 1}^n) } - I({W_1};{Y_{2,i}}|{W_0},{Y_{1}^{i - 1}},Y_{2,i + 1}^n)]  + n\varepsilon^\prime\nonumber\\\label{ghabli1}
&\mathop  = \limits^{(d)} \sum\limits_{i = 1}^n {[I({V_{1,i}};{Y_{1,i}}|{U_i}) - I({V_{1,i}};{Y_{2,i}}|{U_i})]}  + n\varepsilon^\prime
\end{align}
\fi
\makeatother
where $(a)$ and $(b)$ result from Fano's inequality in (\ref{fano1}) and (\ref{fano2}). The equality $(c)$ results from \cite[Lemma~17.12]{csiszarbook} (i.e., Csisz\'{a}r's sum Lemma \cite{CsiszarKorner}), and setting $\varepsilon^\prime={\varepsilon _1} + {\varepsilon _2}$.
The equality $(d)$ results from the following definitions of the random variables in (\ref{Ui})-(\ref{V2i}).
\begin{align}
\label{Ui}
&{U_i} = {W_0},{Y_{1}^{i - 1}},Y_{2,i + 1}^n\\
\label{V1i}
&{V_{1,i}} = ({W_1},{U_i})\\
\label{V2i}
&{V_{2,i}} = ({W_2},{U_i}).
\end{align}
Now, we have
\noindent
\makeatletter%
\if@twocolumn%
\begin{align}
&H({W_1}|{Y_{2}^n})
%\le n\sum\limits_{i = 1}^n {\frac{1}{n}[I({V_{1,Q}};{Y_{1,Q}}|{U_Q}) - } I({V_{1,Q}};{Y_{2,Q}}|{U_Q})]\nonumber\\
%&\,\,\,\,\,\,\,\,\,\,\,\,\,\,\,\,\,\,\,\,\,\,\,\,\,\,\,\,\,\,\,\,\,\,\,\,\,\,\,\,\,\,\,\,\,\,\,\,\,\,\,\,\,\,\,\,\,\,\,\,\,\,\,\,\,\,\,\,\,\,\,\,\,\,\,\,\,\,\,\,\,\,\,\,\,\,\,\,\,\,\,\,\,\,\,\,\,\,\,\,\,\,\,\,\,\,\,\,\,\,\,\,\,\,\,\,\,\,\,\,\,\,\,\,\,\,\,\,\,\,\,\,\,\,\,\,\,\,
%+n\varepsilon^\prime\nonumber\\
\mathop  \le \limits^{(a)} n\sum\limits_{i = 1}^n {p(Q = i)[I({V_{1,Q}};{Y_{1,Q}}|{U_Q},Q = i)  }\nonumber\\
&\,\,\,\,- I({V_{1,Q}};{Y_{2,Q}}|{U_Q},Q = i)]  + n\varepsilon^\prime\nonumber\\
& = n[I({V_{1,Q}};{Y_{1,Q}}|{U_Q},Q) - I({V_{1,Q}};{Y_{2,Q}}|{U_Q},Q)] + n\varepsilon^\prime\nonumber\\ \label{NerkhR1}
&\mathop  = \limits^{(b)} n[I({V_1};Y_{1}|U) - I({V_1};Y_{2}|U)] + n\varepsilon^\prime
\end{align}
\else
\begin{align}
H({W_1}|{Y_{2}^n})
%\le n\sum\limits_{i = 1}^n {\frac{1}{n}[I({V_{1,Q}};{Y_{1,Q}}|{U_Q}) - } I({V_{1,Q}};{Y_{2,Q}}|{U_Q})]\nonumber\\
%&\,\,\,\,\,\,\,\,\,\,\,\,\,\,\,\,\,\,\,\,\,\,\,\,\,\,\,\,\,\,\,\,\,\,\,\,\,\,\,\,\,\,\,\,\,\,\,\,\,\,\,\,\,\,\,\,\,\,\,\,\,\,\,\,\,\,\,\,\,\,\,\,\,\,\,\,\,\,\,\,\,\,\,\,\,\,\,\,\,\,\,\,\,\,\,\,\,\,\,\,\,\,\,\,\,\,\,\,\,\,\,\,\,\,\,\,\,\,\,\,\,\,\,\,\,\,\,\,\,\,\,\,\,\,\,\,\,\,
%+n\varepsilon^\prime\nonumber\\
&\mathop  \le \limits^{(a)} n\sum\limits_{i = 1}^n {p(Q = i)[I({V_{1,Q}};{Y_{1,Q}}|{U_Q},Q = i)  }- I({V_{1,Q}};{Y_{2,Q}}|{U_Q},Q = i)]  + n\varepsilon^\prime\nonumber\\
& = n[I({V_{1,Q}};{Y_{1,Q}}|{U_Q},Q) - I({V_{1,Q}};{Y_{2,Q}}|{U_Q},Q)] + n\varepsilon^\prime\nonumber\\ \label{NerkhR1}
&\mathop  = \limits^{(b)} n[I({V_1};Y_{1}|U) - I({V_1};Y_{2}|U)] + n\varepsilon^\prime
\end{align}
\fi
\makeatother
where (a) results from considering $Q$ with a uniform distribution over $\{ 1,2,...,n\}$ outcomes and (b) is due to defining ${V_{1,Q}} = {V_1},\,{V_{2,Q}} = {V_2},\,{Y_{1,Q}} = Y_{1},\,{Y_{2,Q}} = Y_{2}$ and $({U_Q},Q) = U$. Using (\ref{Privatesecrecy}) and (\ref{NerkhR1}) we derive the bound on $R_{1}$ as follows

\begin{equation}\label{NerxR1}
R_{1} \leq I({V_1};Y_{1}|U) - I({V_1};Y_{2}|U).
\end{equation}
%%%%%%%%%%%%%%%%%%%%%%%%%%%%%

Now, we derive the bound on $R_2$. Using (\ref{Privatesecrecy}) and proceeding the same way as for deriving the bound on $H(W_1|Y_2^n)$, the bound on  $H(W_2|Y_2^n)$ and hence the bound on $R_2$ can be derived as follows:
%%%%%%%%%%%%%%%%%%%
\begin{equation}\label{NerxR2}
R_{2} \leq I({V_2};Y_{1}|U) - I({V_2};Y_{2}|U)
\end{equation}

Based on (\ref{TarifSecrecy}) we have: for any $\varepsilon> 0$,
\begin{equation}
\label{security3}
  n({R_1}+{R_2}) - n\varepsilon  \le H({W_1},{W_2}|{Y_{2}^n})
\end{equation}for all sufficiently large $n$. Hence, to derive the bound on $R_1+R_2$ we first derive the bound on $H(W_1,W_2|Y_2^n)$ as follows:
\noindent
\makeatletter%
\if@twocolumn%
\begin{align}
&H({W_1},{W_2}|{Y_{2}^n}) = H({W_1},{W_2}|{Y_{2}^n},{W_0}) + I({W_1},{W_2};{W_0}|{Y_{2}^n})\nonumber\\
& \mathop  \le \limits^{(a)} H({W_1},{W_2}|{Y_{2}^n},{W_0}) + n{\varepsilon _2}\nonumber\\
& \mathop  \le \limits^{(b)} H({W_1},{W_2}|{Y_{2}^n},{W_0}) - H({W_1},{W_2}|{Y_{1}^n},{W_0}) + n({\varepsilon _1} + {\varepsilon _2})\nonumber\\
& = I({W_1},{W_2};{Y_{1}^n}|{W_0}) - I({W_1},{W_2};{Y_{2}^n}|{W_0}) + n({\varepsilon _1} + {\varepsilon _2})\nonumber\\
& = \sum\limits_{i = 1}^n {[I({W_1},{W_2};{Y_{1,i}}|{W_0},{Y_{1}^{i - 1}})}\nonumber\\
&\,\,\,\,- I({W_1},{W_2};{Y_{2,i}}|{W_0},Y_{2,i + 1}^n)] + n({\varepsilon _1} + {\varepsilon _2})\nonumber\\
%\end{align}
& \mathop  = \limits^{(c)} \sum\limits_{i = 1}^n {[I({W_1},{W_2};{Y_{1,i}}|{W_0},{Y_{1}^{i - 1}},Y_{2,i + 1}^n)} \nonumber\\
&\,\,\,- I({W_1},{W_2};{Y_{2,i}}|{W_0},{Y_{1}^{i - 1}},Y_{2,i + 1}^n)]  + n\varepsilon^\prime\nonumber\\\label{sumR1R222}
& \mathop  = \limits^{(d)} \sum\limits_{i = 1}^n {[I({V_{1,i}},{V_{2,i}};{Y_{1,i}}|{U_i}) - I({V_{1,i}},{V_{2,i}};{Y_{2,i}}|{U_i})]}  + n\varepsilon^\prime
%& = n\sum\limits_{i = 1}^n {\frac{1}{n}[I({V_{1,i}},{V_{2,i}};{Y_{1,i}}|{U_i}) - I({V_{1,i}},{V_{2,i}};{Y_{2,i}}|{U_i})]}  + n({\varepsilon _1} + {\varepsilon _2}),
\end{align}
\else
\begin{align}
H({W_1},{W_2}|{Y_{2}^n}) &= H({W_1},{W_2}|{Y_{2}^n},{W_0}) + I({W_1},{W_2};{W_0}|{Y_{2}^n})\nonumber\\
& \mathop  \le \limits^{(a)} H({W_1},{W_2}|{Y_{2}^n},{W_0}) + n{\varepsilon _2}\nonumber\\
& \mathop  \le \limits^{(b)} H({W_1},{W_2}|{Y_{2}^n},{W_0}) - H({W_1},{W_2}|{Y_{1}^n},{W_0}) + n({\varepsilon _1} + {\varepsilon _2})\nonumber\\
& = I({W_1},{W_2};{Y_{1}^n}|{W_0}) - I({W_1},{W_2};{Y_{2}^n}|{W_0}) + n({\varepsilon _1} + {\varepsilon _2})\nonumber\\
& = \sum\limits_{i = 1}^n {[I({W_1},{W_2};{Y_{1,i}}|{W_0},{Y_{1}^{i - 1}})} - I({W_1},{W_2};{Y_{2,i}}|{W_0},Y_{2,i + 1}^n)]  + n({\varepsilon _1} + {\varepsilon _2})\nonumber\\
%\end{align}
& \mathop  = \limits^{(c)} \sum\limits_{i = 1}^n {[I({W_1},{W_2};{Y_{1,i}}|{W_0},{Y_{1}^{i - 1}},Y_{2,i + 1}^n)} \nonumber\\
&\,\,\,- I({W_1},{W_2};{Y_{2,i}}|{W_0},{Y_{1}^{i - 1}},Y_{2,i + 1}^n)]  + n\varepsilon^\prime\nonumber\\\label{sumR1R222}
& \mathop  = \limits^{(d)} \sum\limits_{i = 1}^n {[I({V_{1,i}},{V_{2,i}};{Y_{1,i}}|{U_i}) - I({V_{1,i}},{V_{2,i}};{Y_{2,i}}|{U_i})]}  + n\varepsilon^\prime
\end{align}
\fi
\makeatother
where $(a)$ and $(b)$ result from Fano's inequality in (\ref{fano1}) and
(\ref{fano2}) respectively. The equality $(c)$ results from Csisz\'{a}r's sum Lemma, and setting $\varepsilon^\prime={\varepsilon _1} + {\varepsilon _2}$.
The equality $(d)$ results from the definitions of the random variables as (\ref{Ui})-(\ref{V2i}). Using (\ref{security3}) and (\ref{sumR1R222}) and by applying the same time-sharing strategy as before, we have
\begin{equation}\label{NerkhejameR1R2}
  {R_1} + {R_2} \leq I({V_1},{V_2};Y_{1}|U) - I({V_1},{V_2};Y_{2}|U)+\varepsilon^\star
\end{equation}where $\varepsilon^\star=\varepsilon + \varepsilon^\prime$.
Finally, we derive the bound on ${R_0}$ as follows:
  \noindent
  \makeatletter%
  \if@twocolumn%
  \begin{align}
n{R_0} &= H({W_0})\nonumber \\
&= I(W_{0};{Y_{1}^n}) + H(W_{0}|{Y_{1}^n}) \nonumber \\
& \le I({W_0};{Y_{1}^n}) + n{\varepsilon_1} \nonumber \\
&= \sum\limits_{i = 1}^n {I({W_0};{Y_{1,i}}|Y_{1}^{i - 1})}  + n{\varepsilon_1} \nonumber \\
&= \sum\limits_{i = 1}^n {[I({W_0},Y_{1}^{i - 1};{Y_{1,i}}) - I(Y_{1}^{i - 1};{Y_{1,i}})]}  + n{\varepsilon_1}. \nonumber
\end{align}
\else
\begin{align}
n{R_0} &= H({W_0})\nonumber \\
&= I(W_{0};{Y_{1}^n}) + H(W_{0}|{Y_{1}^n}) \nonumber \\
& \le I({W_0};{Y_{1}^n}) + n{\varepsilon_1} \nonumber \\
&= \sum\limits_{i = 1}^n {I({W_0};{Y_{1,i}}|Y_{1}^{i - 1})}  + n{\varepsilon_1} \nonumber %\\
\end{align}
\begin{align}
&= \sum\limits_{i = 1}^n {[I({W_0},Y_{1}^{i - 1};{Y_{1,i}}) - I(Y_{1}^{i - 1};{Y_{1,i}})]}  + n{\varepsilon_1}. \nonumber
\end{align}
  \fi
  \makeatother

So, we have
\noindent
\makeatletter%
\if@twocolumn%
\begin{align}
n{R_0} &\le \sum\limits_{i = 1}^n {I({W_0},{Y_{1}^{i - 1}};{Y_{1,i}})}  + n{\varepsilon_1} \nonumber \\
&= \sum\limits_{i = 1}^n {[I({W_0},{Y_{1}^{i - 1}},Y_{2,i + 1}^n;{Y_{1,i}})}\nonumber\\
&\,\,\,\,- I(Y_{2,i + 1}^n;{Y_{1,i}}|{W_0},{Y_{1}^{i - 1}})] + n{\varepsilon_1} \nonumber \\
& \le \sum\limits_{i = 1}^n {I({W_0},{Y_{1}^{i - 1}},Y_{2,i + 1}^n;{Y_{1,i}})}  + n{\varepsilon_1} \nonumber%\\
\end{align}
\begin{align}
\label{outerR0}
&= \sum\limits_{i = 1}^n {I({U_i};{Y_{1,i}})  + n{\varepsilon_1}}.
\end{align}
\else
\begin{align}
n{R_0} &\le \sum\limits_{i = 1}^n {I({W_0},{Y_{1}^{i - 1}};{Y_{1,i}})}  + n{\varepsilon_1} \nonumber \\
&= \sum\limits_{i = 1}^n {[I({W_0},{Y_{1}^{i - 1}},Y_{2,i + 1}^n;{Y_{1,i}})}- I(Y_{2,i + 1}^n;{Y_{1,i}}|{W_0},{Y_{1}^{i - 1}})] + n{\varepsilon_1} \nonumber \\
& \le \sum\limits_{i = 1}^n {I({W_0},{Y_{1}^{i - 1}},Y_{2,i + 1}^n;{Y_{1,i}})}  + n{\varepsilon_1} \nonumber \\
\label{outerR0}
&= \sum\limits_{i = 1}^n {I({U_i};{Y_{1,i}})  + n{\varepsilon_1}}.
\end{align}
\fi
\makeatother

By applying the same time-sharing strategy as before, we have
\begin{equation}\label{111}
{R_0} \leq {I(U;Y_{1})} + {\varepsilon_1}.
\end{equation}

Also we have,
%%%%%%%%%%%%%%%%%%%%%%%%%%%%%%%%%%%%%%%%%%%%%%%%%%%%%%%%%%%%%%
\begin{align}
n{R_0} &= H({W_0})\nonumber \\
&= I(W_{0};{Y_{2}^n}) + H(W_{0}|{Y_{2}^n}) \nonumber \\
& \le I({W_0};{Y_{2}^n}) + n{\varepsilon_2} \nonumber \\
&= \sum\limits_{i = 1}^n {I({W_0};{Y_{2,i}}|Y_{2,i+1}^{n})}  + n{\varepsilon_2} \nonumber \\
&= \sum\limits_{i = 1}^n {[I({W_0},Y_{2,i+1}^{n};{Y_{2,i}}) - I(Y_{2,i+1}^{n};{Y_{2,i}})]}  + n{\varepsilon_2}. \nonumber
\end{align}

So, we have
\noindent
\makeatletter%
\if@twocolumn%
\begin{align}
n{R_0} &\le \sum\limits_{i = 1}^n {I({W_0},Y_{2,i+1}^{n};{Y_{2,i}})}  + n{\varepsilon_2} \nonumber \\
&= \sum\limits_{i = 1}^n {[I({W_0},{Y_{1}^{i - 1}},Y_{2,i + 1}^n;{Y_{2,i}})}\nonumber\\
&\,\,\,\, - I({Y_{1}^{i - 1}};{Y_{2,i}}|{W_0},Y_{2,i + 1}^n)] + n{\varepsilon_2} \nonumber \\
& \le \sum\limits_{i = 1}^n {I({W_0},{Y_{1}^{i - 1}},Y_{2,i + 1}^n;{Y_{2,i}})}  + n{\varepsilon_2} \nonumber\\
\label{outerR0dov}
&= \sum\limits_{i = 1}^n {I({U_i};{Y_{2,i}})  + n{\varepsilon_1}}.
\end{align}
\else
\begin{align}
n{R_0} &\le \sum\limits_{i = 1}^n {I({W_0},Y_{2,i+1}^{n};{Y_{2,i}})}  + n{\varepsilon_2} \nonumber \\
&= \sum\limits_{i = 1}^n {[I({W_0},{Y_{1}^{i - 1}},Y_{2,i + 1}^n;{Y_{2,i}})}- I({Y_{1}^{i - 1}};{Y_{2,i}}|{W_0},Y_{2,i + 1}^n)] + n{\varepsilon_2} \nonumber \\
& \le \sum\limits_{i = 1}^n {I({W_0},{Y_{1}^{i - 1}},Y_{2,i + 1}^n;{Y_{2,i}})}  + n{\varepsilon_2} \nonumber \\\label{outerR0dov}
&= \sum\limits_{i = 1}^n {I({U_i};{Y_{2,i}})  + n{\varepsilon_2}}.
\end{align}
\fi
\makeatother

By applying the same time-sharing strategy as before, we have
\begin{equation}\label{111111}
{R_0} \leq {I(U;Y_{2})} + {\varepsilon_2}.
\end{equation}
%%%%%%%%%%%%%%%%%%%%%%%%%%%%%%%%%%%%%%%%%%%%%%%%%%%%%%%%%%%%%%

Therefore, from (\ref{111}) and (\ref{111111}) we have:
\begin{equation}\label{333}
{R_0} \leq \min \{I({U};Y_{1}),I({U};Y_{2})\}.
\end{equation}

Considering (\ref{NerxR1}), (\ref{NerxR2}), (\ref{NerkhejameR1R2}) and (\ref{333}), the region in (\ref{R0})-(\ref{jameR1R2}) is obtained. The bounds on cardinality of $|\cal U|$, $|{\cal V}_1|$ and $|{\cal V}_2|$ can be derived by following the steps in \cite[Appendix]{CsiszarKorner}. This completes the proof.
\section{Proof for Theorem~\ref{achievthm}}\label{achievproof}
\begin{figure*}%[ht]
  \centering
  \includegraphics[width=12.50cm]{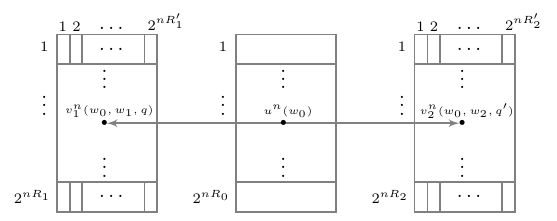}% asli 12.50cm 8.50
  \caption{Coding scheme.}
  \label{divomin}
\end{figure*}

As described in Section~III.B, we use the coding structure as illustrated in Fig.~\ref{divomin}. More precisely, the codebook generation is as follows:
\subsubsection{Codebook generation} Fix $p(u),p(v_1|u),p(v_2|u),p({x_1}|v_1)$ and $p({x_2}|v_2)$. Let
\begin{equation}\label{jamesecur}
 {R'_1} + {R'_2} = I({V_1},{V_{_2}};Y_{2}|U) - {\varepsilon},
\end{equation}where ${\varepsilon} > 0$ and ${\varepsilon} \to 0$ as $n \to \infty$.
\setlength{\textfloatsep}{10pt plus 1.0pt minus 2.0pt}
\begin{enumerate}[i)]
\item Generate ${2^{n{R_0}}}$ length-$n$ codewords ${u^n}$ through $p({u^n}) = \prod\nolimits_{i = 1}^n {p({u_i})}$ and index them as ${u^n}({w_0}),\,\,\,{w_0} \in \{ 1,...,{2^{n{R_0}}}\} $.
\item For each codeword ${u^n}({w_0})$, generate ${2^{n{{\widetilde R}_1}}}$ length-$n$ codewords $v_{1}^n$ through $p(v_1^n|{u^n}) = \prod\nolimits_{i = 1}^n {p({v_{1,i}}|{u_i})}$, where ${\widetilde R_1} = {R_1} + {R'_1}$. Then, randomly bin ${2^{n{{\widetilde R}_1}}}$ codewords into ${2^{n{R_1}}}$ bins, and index them as $v_{1}^n({w_0},{w_1},q)$, where ${w_1} \in \{ 1,...,{2^{n{R_1}}}\}$ is the bin number and ${q} \in \mathcal{Q} = \{ 1,...,{2^{n{R'_1}}}\} $ is the index of codewords in the bin number ${w_1}$.
\end{enumerate}
The codebook for user 2 is generated in the same way.
\subsubsection{Encoding}
Assume that $(w_{0},w_{1})$  is the message pair to be transmitted, the encoder $g_{1}$ randomly chooses index $q$ corresponding to $(w_{0},w_{1})$ and then generates a codeword $X_1^n$ at random according to $\prod\nolimits_{i = 1}^n {p({x_{1,i}}|{v_{1,i}})}$. Transmitter~2 uses the same way to encode $({w_0},{w_2})$.
\subsubsection{Decoding and Probability of error}$\\$
\textbf{Decoding:}
\begin{itemize}
\item The legitimate receiver (Receiver~1) decodes $(\widehat w_{01},\widehat w_{1},\widehat w_{2})$ by looking for the unique $(u^n,v_1^n,v_2^n)$ such that $({u^n}(\widehat w_{01}),v_{1}^n(\widehat w_{01},\widehat w_{1},q),v_2^n(\widehat w_{01},\widehat w_{2},q'),{y_{1}^n}) \in T_{\varepsilon} ^n({P_{UV_1V_2Y_{1}}})$.
\item Receiver~2 decodes ${\hat w_{02}}$ by looking for the unique ${u^n}$ such that $({u^n}(\widehat w_{02}),{y_{2}^n}) \in T_{\varepsilon} ^n({P_{UY_{2}}})$.$\\$
\end{itemize}
\textbf{Probability of Error Analysis:}
%By using joint typicality decoding \cite{ElgamalKim} and considering ${\tilde R_1} = {R_1} + {R'_1}$ and ${\tilde R_2} = {R_2} + {R'_2}$, it can easily be shown that the probability of error goes to zero as $n\rightarrow\infty$ if we choose:
%%%%%%%%%%%%%%%%%%%%% Begin Probability of Error %%%%%%%%%%%%%%%%%%%%%%%%%%%%%%
Define the events
\noindent
\makeatletter%
\if@twocolumn%
%\textcolor{blue}{
\begin{align}
&{E_{r,{w_0},{w_1},{w_2}}} = \{ ({u^n}({\widehat w_{01}}),v_1^n({{\widehat w}_{01}},{{\widehat w}_{11}},q)\nonumber\\
&\,\,\,\, ,v_2^n({{\widehat w}_{01}},{{\widehat w}_{21}},q'),{y_{1}^n}) \in T_{\varepsilon} ^n({P_{UV_1V_2Y_{1}}})\} \\
&{E_{e,{w_0}}} = \{ ({u^n}({{\widehat w}_{02}}),{y_{2}^n}) \in T_{\varepsilon} ^n({P_{UY_{2}}})\}.
\end{align}%}
\else
%\textcolor{blue}{
\begin{align}
&{E_{r,{w_0},{w_1},{w_2}}} = \{ ({u^n}({\widehat w_{01}}),v_1^n({{\widehat w}_{01}},{{\widehat w}_{11}},q),v_2^n({{\widehat w}_{01}},{{\widehat w}_{21}},q'),{y_{1}^n}) \in T_{\varepsilon} ^n({P_{UV_1V_2Y_{1}}})\} \\
&{E_{e,{w_0}}} = \{ ({u^n}({{\widehat w}_{02}}),{y_{2}^n}) \in T_{\varepsilon} ^n({P_{UY_{2}}})\}.
\end{align}%}
\fi
\makeatother
Without loss of generality, we assume that $({w_0},{w_1},{w_2}) = (1,1,1)$ was sent. We can bound the probability of error in Receiver~1 using the union bound:
\noindent
\makeatletter%
\if@twocolumn%
%\textcolor{blue}{
\begin{align}
&P_{e,1}^n(1,1,1) = \mbox{Pr}\{ E_{r,1,1,1}^c\bigcup\limits_{{w_0} \ne 1} {{E_{r,{w_0},1,1}}} \bigcup\limits_{{w_1} \ne 1} {{E_{r,1,{w_1},1}}} \nonumber\\
&\,\,\,\,\bigcup\limits_{{\scriptstyle{{w_0}}} \ne 1\hfill\atop {\scriptstyle{w_1}} \ne 1} {{E_{r,{w_0},w_{1},1}}} \bigcup\limits_{{w_2} \ne 1} {{E_{r,1,1,{w_2}}}} \bigcup\limits_{{\scriptstyle{{w_0}}} \ne 1\hfill\atop {\scriptstyle{w_2}} \ne 1} {{E_{r,{w_0},1,w_{2}}}}\nonumber\\
&\,\,\,\,\bigcup\limits_{\scriptstyle{w_1} \ne 1\hfill\atop
\scriptstyle{w_2} \ne 1\hfill} {{E_{r,1,{w_1},{w_2}}}} \bigcup\limits_{\scriptstyle{w_0} \ne 1\hfill\atop
{\scriptstyle{w_1} \ne 1\hfill\atop
\scriptstyle{w_2} \ne 1\hfill}} {{E_{r,{w_0},{w_1},{w_2}}}} \} \\\nonumber
& \le \mbox{Pr}\{E_{r,1,1,1}^c\} +\sum\limits_{{w_0} \ne 1} \mbox{Pr}\{{E_{r,{w_0},1,1}}\}\nonumber\\
&\,\,\,\,+ \sum\limits_{{w_1} \ne 1} {\sum\limits_q {\mbox{Pr}\{{E_{r,1,{w_1},1}}\}}}  + \sum\limits_{{w_0} \ne 1}\sum\limits_{{w_1} \ne 1} {\sum\limits_q {\mbox{Pr}\{{E_{r,{w_0},{w_1},1}}\}}} \nonumber\\
&\,\,\,\,+  \sum\limits_{{w_2} \ne 1} {\sum\limits_{q'} {\mbox{Pr}\{{E_{r,1,1,{w_2}}}\}} } + \sum\limits_{{w_0} \ne 1}\sum\limits_{{w_2} \ne 1} {\sum\limits_{q'} {\mbox{Pr}\{{E_{r,{w_0},1,{w_2}}}\}}} \nonumber\\
&\,\,\,\, + {\sum\limits_{{w_1} \ne 1} {\sum\limits_q {\sum\limits_{{w_2} \ne 1} {\sum\limits_{q'} {\mbox{Pr}\{{E_{r,1,{w_1},{w_2}}}\}} } } } } \nonumber\\
&\,\,\,\,+\sum\limits_{{w_0} \ne 1} {\sum\limits_{{w_1} \ne 1} {\sum\limits_{q} {\sum\limits_{{w_2} \ne 1} {\sum\limits_{q'} {\mbox{Pr}\{{E_{r,{w_0},{w_1},{w_2}}}\}} } } } }.\label{Perrorr}
\end{align}%}
\else
%\textcolor{blue}{
\begin{align}
&P_{e,1}^n(1,1,1) = \mbox{Pr}\{ E_{r,1,1,1}^c\bigcup\limits_{{w_0} \ne 1} {{E_{r,{w_0},1,1}}} \bigcup\limits_{{w_1} \ne 1} {{E_{r,1,{w_1},1}}} \bigcup\limits_{{\scriptstyle{{w_0}}} \ne 1\hfill\atop {\scriptstyle{w_1}} \ne 1} {{E_{r,{w_0},w_{1},1}}} \bigcup\limits_{{w_2} \ne 1} {{E_{r,1,1,{w_2}}}} \bigcup\limits_{{\scriptstyle{{w_0}}} \ne 1\hfill\atop {\scriptstyle{w_2}} \ne 1} {{E_{r,{w_0},1,w_{2}}}}\nonumber\\
&\,\,\,\,\,\,\bigcup\limits_{\scriptstyle{w_1} \ne 1\hfill\atop
\scriptstyle{w_2} \ne 1\hfill} {{E_{r,1,{w_1},{w_2}}}} \bigcup\limits_{\scriptstyle{w_0} \ne 1\hfill\atop
{\scriptstyle{w_1} \ne 1\hfill\atop
\scriptstyle{w_2} \ne 1\hfill}} {{E_{r,{w_0},{w_1},{w_2}}}} \} \\\nonumber
& \le \mbox{Pr}\{E_{r,1,1,1}^c\} +\sum\limits_{{w_0} \ne 1} \mbox{Pr}\{{E_{r,{w_0},1,1}}\}+ \sum\limits_{{w_1} \ne 1} {\sum\limits_q {\mbox{Pr}\{{E_{r,1,{w_1},1}}\}}}  +\sum\limits_{{w_0} \ne 1}\sum\limits_{{w_1} \ne 1} {\sum\limits_q {\mbox{Pr}\{{E_{r,{w_0},{w_1},1}}\}}} \nonumber\\
&\,\,\,\,\,\,+  \sum\limits_{{w_2} \ne 1} {\sum\limits_{q'} {\mbox{Pr}\{{E_{r,1,1,{w_2}}}\}} } + \sum\limits_{{w_0} \ne 1}\sum\limits_{{w_2} \ne 1} {\sum\limits_{q'} {\mbox{Pr}\{{E_{r,{w_0},1,{w_2}}}\}}} \nonumber\\
&\,\,\,\,\,\, + {\sum\limits_{{w_1} \ne 1} {\sum\limits_q {\sum\limits_{{w_2} \ne 1} {\sum\limits_{q'} {\mbox{Pr}\{{E_{r,1,{w_1},{w_2}}}\}} } } } }+ \sum\limits_{{w_0} \ne 1} {\sum\limits_{{w_1} \ne 1} {\sum\limits_{q} {\sum\limits_{{w_2} \ne 1} {\sum\limits_{q'} {\mbox{Pr}\{{E_{r,{w_0},{w_1},{w_2}}}\}} } } } }.\label{Perrorr}
\end{align}%}
\fi
\makeatother
In the same way we can bound the probability of error in Receiver~2 using the union bound as:
\begin{align}
P_{e,2}^n(1,1,1) &= \mbox{Pr}\{{E_{e,1}^c} \bigcup\limits_{{w_0} \ne 1} {{E_{e,{w_0}}}}\}\nonumber\\\label{Perrore}
&\leq \mbox{Pr}\{E_{e,1}^c\} + \sum\limits_{{w_0} \ne 1} \mbox{Pr}\{{{E_{e,{w_0}}}}\}
\end{align}
From the Asymptotic Equipartition Property (AEP)\cite[Chapter~3]{CoverThomas} and \cite[Thm. 15.2.1, 15.2.3]{CoverThomas}, it follows that
\noindent
\makeatletter%
\if@twocolumn%
%\textcolor{blue}{
\begin{align}
&\mbox{Pr}\{E_{r,1,1,1}^c\} \le \varepsilon \\
&\mbox{Pr}\{{E_{r,{w_0},1,1}}\} \le {2^{ - n[I(U,{V_1},{V_2};Y_{1}) - \varepsilon ]}}\\
&\mbox{Pr}\{{E_{r,1,{w_1},1}}\} \le {2^{ - n[I({V_1};Y_{1}|{V_2},U) - \varepsilon ]}}\\
&\mbox{Pr}\{{E_{r,{w_0},{w_1},1}}\} \le {2^{ - n[I(U,{V_1},{V_2};Y_{1}) - \varepsilon ]}}\\
&\mbox{Pr}\{{E_{r,1,1,{w_2}}}\} \le {2^{ - n[I({V_2};Y_{1}|{V_1},U) - \varepsilon ]}}\\
&\mbox{Pr}\{{E_{r,{w_0},1,{w_2}}}\} \le {2^{ - n[I(U,{V_1},{V_2};Y_{1}) - \varepsilon ]}}\\
&\mbox{Pr}\{{E_{r,1,{w_1},{w_2}}}\} \le {2^{ - n[I({V_1},{V_2};Y_{1}|U) - \varepsilon ]}}\\
&\mbox{Pr}\{{E_{r,{w_0},{w_1},{w_2}}}\} \le {2^{ - n[I(U,{V_1},{V_2};Y_{1}) - \varepsilon ]}}\\
%\end{align}
%\begin{align}
&\mbox{Pr}\{E_{e,1}^c\} \le \varepsilon \\
&\mbox{Pr}\{{E_{e,{w_0}}}\} \le {2^{ - n[I(U;Y_{2}) - \varepsilon ]}}
\end{align}%}
\else
%\textcolor{blue}{
\begin{align}
&\mbox{Pr}\{E_{r,1,1,1}^c\} \le \varepsilon \\
&\mbox{Pr}\{{E_{r,{w_0},1,1}}\} \le {2^{ - n[I(U,{V_1},{V_2};Y_{1}) - \varepsilon ]}}\\
&\mbox{Pr}\{{E_{r,1,{w_1},1}}\} \le {2^{ - n[I({V_1};Y_{1}|{V_2},U) - \varepsilon ]}}\\
&\mbox{Pr}\{{E_{r,{w_0},{w_1},1}}\} \le {2^{ - n[I(U,{V_1},{V_2};Y_{1}) - \varepsilon ]}}\\
&\mbox{Pr}\{{E_{r,1,1,{w_2}}}\} \le {2^{ - n[I({V_2};Y_{1}|{V_1},U) - \varepsilon ]}}\\
&\mbox{Pr}\{{E_{r,{w_0},1,{w_2}}}\} \le {2^{ - n[I(U,{V_1},{V_2};Y_{1}) - \varepsilon ]}}\\
&\mbox{Pr}\{{E_{r,1,{w_1},{w_2}}}\} \le {2^{ - n[I({V_1},{V_2};Y_{1}|U) - \varepsilon ]}}\\
&\mbox{Pr}\{{E_{r,{w_0},{w_1},{w_2}}}\} \le {2^{ - n[I(U,{V_1},{V_2};Y_{1}) - \varepsilon ]}}\\
&\mbox{Pr}\{E_{e,1}^c\} \le \varepsilon \\
&\mbox{Pr}\{{E_{e,{w_0}}}\} \le {2^{ - n[I(U;Y_{2}) - \varepsilon ]}}
\end{align}%}
\fi
\makeatother
where ${\varepsilon} > 0$ and ${\varepsilon} \to 0$ as $n \to \infty$. Hence, (\ref{Perrorr}) and (\ref{Perrore}) are respectively bounded by
\noindent
\makeatletter%
\if@twocolumn%
%\textcolor{blue}{
\begin{align}
&P_{e,1}^n(1,1,1) \le \varepsilon  +{2^{n{{R}_0}}} \times {2^{ - n[I(U,{V_1},{V_2};Y_{1})- \varepsilon ]}} \nonumber\\
&\,\,\,\, + {2^{n{{\tilde R}_1}}} \times {2^{ - n[I({V_1};Y_{1}|{V_2},U) - \varepsilon ]}} \nonumber\\
&\,\,\,\, + {2^{n{{R}_0}+{\tilde R}_1}} \times {2^{ - n[I(U,{V_1},{V_2};Y_{1})- \varepsilon ]}} \nonumber\\
&\,\,\,\, + {2^{n{{\tilde R}_2}}} \times {2^{ - n[I({V_2};Y_{1}|{V_1},U) - \varepsilon ]}}\nonumber\\
&\,\,\,\, + {2^{n{{R}_0}+{\tilde R}_2}} \times {2^{ - n[I(U,{V_1},{V_2};Y_{1})- \varepsilon ]}}\nonumber\\\label{Perror2}
&\,\,\,\, + {2^{n({{\tilde R}_1} + {{\tilde R}_2})}} \times {2^{ - n[I({V_1},{V_2};Y_{1}|U) - \varepsilon ]}} \nonumber\\
&\,\,\,\, + {2^{n({R_0} + {{\tilde R}_1} + {{\tilde R}_2})}} \times {2^{ - n[I(U,{V_1},{V_2};Y_{1}) - \varepsilon ]}}
\end{align}%}
\else
%\textcolor{blue}{
\begin{align}
&P_{e,1}^n(1,1,1) \le \varepsilon  +{2^{n{{R}_0}}} \times {2^{ - n[I(U,{V_1},{V_2};Y_{1})- \varepsilon ]}} + {2^{n{{\tilde R}_1}}} \times {2^{ - n[I({V_1};Y_{1}|{V_2},U) - \varepsilon ]}} \nonumber\\
&\,\,\,+{2^{n{{R}_0}+{\tilde R}_1}} \times {2^{ - n[I(U,{V_1},{V_2};Y_{1})- \varepsilon ]}} + {2^{n{{\tilde R}_2}}} \times {2^{ - n[I({V_2};Y_{1}|{V_1},U) - \varepsilon ]}}+{2^{n{{R}_0}+{\tilde R}_2}} \times {2^{ - n[I(U,{V_1},{V_2};Y_{1})- \varepsilon ]}}\nonumber\\\label{Perror2}
&\,\,\,+ {2^{n({{\tilde R}_1} + {{\tilde R}_2})}} \times {2^{ - n[I({V_1},{V_2};Y_{1}|U) - \varepsilon ]}} + {2^{n({R_0} + {{\tilde R}_1} + {{\tilde R}_2})}} \times {2^{ - n[I(U,{V_1},{V_2};Y_{1}) - \varepsilon ]}}
\end{align}%}
\fi
\makeatother
and
\begin{align}\label{Perror3}
P_{e,2}^n(1) \le \varepsilon  + {2^{n{R_0}}} \times {2^{ - n[I(U;Y_{2}) - \varepsilon ]}}.
\end{align}Due to (\ref{Perror3}) and (\ref{Perror2}), to generate $P_e^n \rightarrow 0$ as $n \rightarrow \infty$, we must choose
\noindent
\makeatletter%
\if@twocolumn%
%\textcolor{blue}{
\begin{align}
\label{r0aval}
{R_0} &\le I(U;Y_{2})\\
\label{r1tildaval}
{{\tilde R}_1} &\le I({V_1};Y_{1}|{V_2},U)\\
\label{r2tildaval}
{{\tilde R}_2} &\le I({V_2};Y_{1}|{V_1},U)\\
\label{r1tildavalvar2tildaval}
{{\tilde R}_1} + {{\tilde R}_2} &\le I({V_1},{V_2};Y_{1}|U)\,\\
\label{majmooaval}
{R_0} + {{\tilde R}_1} + {{\tilde R}_2} &\le I(U,{V_1},{V_2};Y_{1}),
\end{align}and
\begin{align}
{R_0} &\le I(U,{V_1},{V_2};Y_{1})\label{ezafir0}\\
{R_0}+{{\tilde R}_1} &\le I(U,{V_1},{V_2};Y_{1})\label{ezafir0r1}\\
{R_0}+{{\tilde R}_2} &\le I(U,{V_1},{V_2};Y_{1})\label{ezafir0r2}
\end{align}%}
\else
%\textcolor{blue}{
\begin{align}
\label{r0aval}
{R_0} &\le I(U;Y_{2})\\
\label{r1tildaval}
{{\tilde R}_1} &\le I({V_1};Y_{1}|{V_2},U)\\
\label{r2tildaval}
{{\tilde R}_2} &\le I({V_2};Y_{1}|{V_1},U)\\
%\end{align}
%\begin{align}
\label{r1tildavalvar2tildaval}
{{\tilde R}_1} + {{\tilde R}_2} &\le I({V_1},{V_2};Y_{1}|U)\,\\
\label{majmooaval}
{R_0} + {{\tilde R}_1} + {{\tilde R}_2} &\le I(U,{V_1},{V_2};Y_{1}),
\end{align}and
\begin{align}
{R_0} &\le I(U,{V_1},{V_2};Y_{1})\label{ezafir0}\\
{R_0}+{{\tilde R}_1} &\le I(U,{V_1},{V_2};Y_{1})\label{ezafir0r1}\\
{R_0}+{{\tilde R}_2} &\le I(U,{V_1},{V_2};Y_{1})\label{ezafir0r2}
\end{align}%}
\fi
\makeatother
then $P_{e,1}^n \leq \varepsilon$ and $P_{e,2}^n \leq \varepsilon$, and from (\ref{Pe}) we have: $P_{e}^n \leq \varepsilon$. Note that in (\ref{r0aval})-(\ref{ezafir0r2}), we have these Markov chains $V_{1}-U-V_{2}$ and $U-(V_{1},V_{2})-(Y_{1},Y_{2})$, and the bounds
(\ref{ezafir0})-(\ref{ezafir0r2}) are redundant because of (\ref{majmooaval}). Therefore, we need to consider only (\ref{r0aval})-(\ref{majmooaval}). Also ${\tilde R_1} = {R_1} + {R'_1}$ and ${\tilde R_2} = {R_2} + {R'_2}$ that by replacing these into (\ref{r0aval})-(\ref{majmooaval}) we have
%%%%%%%%%%%%%%%%%%%%% End Probability of Error %%%%%%%%%%%%%%%%%%%%%%%%%%%%%%
\begin{align}
\label{r0dovom}
{R_0} &\le I(U;Y_{2})\\
\label{r1varprim1dovom}
{R_1} + {R'_1} &\le I({V_1};Y_{1}|{V_2},U)\\
\label{r2varprim2dovom}
{R_2} + {R'_2} &\le I({V_2};Y_{1}|{V_1},U)\\
\label{r1varprim1dovomr2varprim2dovom}
{R_1} + {R'_1} + {R_2} + {R'_2} &\le I({V_1},{V_2};Y_{1}|U)\,\\
\label{majmoodovom}
{R_0} + {R_1} + {R'_1} + {R_2} + {R'_2} &\le I(U,{V_1},{V_2};Y_{1})
\end{align}
\subsubsection{Equivocation computation}
\noindent
\makeatletter%
\if@twocolumn%
\begin{align}
&H({W_1},{W_2}|{Y_{2}^n}) \ge H({W_1},{W_2}|{Y_{2}^n},{U^n})\nonumber\\
&= H({W_1},{W_2},{Y_{2}^n}|{U^n}) - H({Y_{2}^n}|{U^n})\nonumber\\
&= H({W_1},{W_2},{Y_{2}^n},V_1^n,V_2^n|{U^n}) \nonumber\\
&\,\,\,\, - H(V_1^n,V_2^n|{W_1},{W_2},{Y_{2}^n},{U^n})- H({Y_{2}^n}|{U^n})\nonumber\\
&= H({W_1},{W_2},V_1^n,V_2^n|{U^n}) + H({Y_{2}^n}|{W_1},{W_2},V_1^n,V_2^n,{U^n})\nonumber\\
&\,\,\,- H(V_1^n,V_2^n|{W_1},{W_2},{Y_{2}^n},{U^n}) - H({Y_{2}^n}|{U^n})\nonumber\\
&\mathop  \ge \limits^{(a)} H(V_1^n,V_2^n|{U^n}) - H(V_1^n,V_2^n|{W_1},{W_2},{Y_{2}^n},{U^n}) \nonumber\\
&\,\,\,+ H({Y_{2}^n}|V_1^n,V_2^n,{U^n}) - H({Y_{2}^n}|{U^n})\nonumber%\\
\end{align}
\begin{align}
\label{Equivocation}
&= H(V_1^n,V_2^n|{U^n}) - H(V_1^n,V_2^n|{W_1},{W_2},{Y_{2}^n},{U^n})\nonumber\\
&\,\,\,\, - I(V_1^n,V_2^n;{Y_{2}^n}|{U^n})
\end{align}
\else
\begin{align}
H({W_1},{W_2}|{Y_{2}^n}) &\ge H({W_1},{W_2}|{Y_{2}^n},{U^n})\nonumber\\
&= H({W_1},{W_2},{Y_{2}^n}|{U^n}) - H({Y_{2}^n}|{U^n})\nonumber\\
&= H({W_1},{W_2},{Y_{2}^n},V_1^n,V_2^n|{U^n})  - H(V_1^n,V_2^n|{W_1},{W_2},{Y_{2}^n},{U^n})- H({Y_{2}^n}|{U^n})\nonumber\\
&= H({W_1},{W_2},V_1^n,V_2^n|{U^n}) + H({Y_{2}^n}|{W_1},{W_2},V_1^n,V_2^n,{U^n})\nonumber\\
&\,\,\,- H(V_1^n,V_2^n|{W_1},{W_2},{Y_{2}^n},{U^n}) - H({Y_{2}^n}|{U^n})\nonumber\\
&\mathop  \ge \limits^{(a)} H(V_1^n,V_2^n|{U^n}) - H(V_1^n,V_2^n|{W_1},{W_2},{Y_{2}^n},{U^n}) \nonumber\\
&\,\,\,+ H({Y_{2}^n}|V_1^n,V_2^n,{U^n}) - H({Y_{2}^n}|{U^n})\nonumber\\\label{Equivocation}
%\end{align}
%\begin{align}
&= H(V_1^n,V_2^n|{U^n}) - H(V_1^n,V_2^n|{W_1},{W_2},{Y_{2}^n},{U^n})- I(V_1^n,V_2^n;{Y_{2}^n}|{U^n})
\end{align}
\fi
\makeatother
where (a) is due to $I(Y_2^n;W_{1},W_{2}|V_1^n,V_2^n,U^n)=0$.

The first term in (\ref{Equivocation}) is given by:
\begin{equation}\label{termaval}
H(V_1^n,V_2^n|{U^n}) = n{\widetilde R_1} + n{\widetilde R_2} = n({R_1} + {R'_1} + {R_2} + {R'_2}).
\end{equation}

We then show that the second term in (\ref{Equivocation}) can be bounded by $H(V_1^n,V_2^n|{W_1},{W_2},{Y_{2}^n},{U^n}) \le n{\varepsilon_1}$, as $n \to \infty $ then ${\varepsilon_1} \to 0$. To this aim, it can be noted that given the message $(W_{1},W_{2})=(w_{1},w_{2})$ and assuming that receiver~2 knows the sequence $U^{n}=u^{n}$, it can decode $(q,q')$ with small probability of error if
\begin{align}
\label{rprimaval}
{R'_1} &\le I({V_1};Y_{2}|{V_2},U)\\
\label{rprimdovom}
{R'_2} &\le I({V_2};Y_{2}|{V_1},U)\\
\label{rprimsevom}
{R'_1} + {R'_2} &\le I({V_1},{V_2};Y_{2}|U).
\end{align}
for sufficiently large $n$.

Using Fano's inequality  implies that $H(V_1^n,V_2^n|{W_1} = {w_1},{W_2} = {w_2},{Y_{2}^n},{U^n}) \le n{\varepsilon_1}$. Hence,
\noindent
\makeatletter%
\if@twocolumn%
\begin{align}\label{esbatedovom}
&H(V_1^n,V_2^n|{W_1},{W_2},Y_{2}^n,{U^n}) = \nonumber\\
&\,\,\,\sum\limits_{w_1} {\sum\limits_{{w_2}} {p({W_1} = {w_1})p({W_2} = {w_2})} }\nonumber\\
&\,\,\,\,\times H(V_1^n,V_2^n|{W_1} = {w_1},{W_2} = {w_2},{Y_{2}^n},{U^n}) \le n{\varepsilon_1}.
\end{align}
\else
\begin{align}\label{esbatedovom}
&H(V_1^n,V_2^n|{W_1},{W_2},Y_{2}^n,{U^n}) = \nonumber\\
&\,\,\,\sum\limits_{w_1} {\sum\limits_{{w_2}} {p({W_1} = {w_1})p({W_2} = {w_2})} }\times H(V_1^n,V_2^n|{W_1} = {w_1},{W_2} = {w_2},{Y_{2}^n},{U^n}) \le n{\varepsilon_1}.
\end{align}
\fi
\makeatother
The last term in (\ref{Equivocation}) is bounded as:
\begin{equation}\label{esbatesevom}
I(V_1^n,V_2^n;{Y_{2}^n}|{U^n})\, \le nI({V_1},{V_2};Y_{2}|U) + n{\varepsilon_2},
\end{equation}as $n \to \infty$ then ${\varepsilon_2} \to 0$ similar to \cite[Lemma 1]{Wyner}. By replacing (\ref{termaval}), (\ref{esbatedovom}) and (\ref{esbatesevom}) in (\ref{Equivocation}) we have:
\noindent
\makeatletter%
\if@twocolumn%
\begin{align}
&H({W_1},{W_2}|{Y_{2}^n}) \ge n({R_1} + {R'_1} + {R_2} + {R'_2}) - n{\varepsilon _1} \nonumber\\
&\,\,\,\,- nI({V_1},{V_2};Y_{2}|U) - n{\varepsilon _2}\\
&= n({R_1} + {R_2} + {R'_1} + {R'_2} - I({V_1},{V_2};Y_{2}|U)) - n\delta \\
&= n({R_1} + {R_2}) - 2n\delta
\end{align}
\else
\begin{align}
H({W_1},{W_2}|{Y_{2}^n}) &\ge n({R_1} + {R'_1} + {R_2} + {R'_2}) - n{\varepsilon _1} - nI({V_1},{V_2};Y_{2}|U) - n{\varepsilon _2}\\
&= n({R_1} + {R_2} + {R'_1} + {R'_2} - I({V_1},{V_2};Y_{2}|U)) - n\delta \\
&= n({R_1} + {R_2}) - 2n\delta
\end{align}
\fi
\makeatother
where $\delta  = {\varepsilon _1} + {\varepsilon _2}$. Finally, by using the Fourier-Motzkin procedure \cite{ElgamalKim} to eliminate ${R'_1}$ and ${R'_2}$ in (\ref{jamesecur}), (\ref{r0dovom})-(\ref{majmoodovom}) and (\ref{rprimaval})-(\ref{rprimsevom}) we obtain the five inequalities in Theorem~\ref{achievthm}. The bounds on cardinality of $|\cal U|$, $|{\cal V}_1|$ and $|{\cal V}_2|$ can be derived by following the steps in \cite[Appendix]{CsiszarKorner}. This completes the proof of Theorem~\ref{achievthm}.
\section{Proof for Theorem~\ref{scthm}}\label{SCproof}
To prove this theorem, we concentrate on outer bound (Theorem~\ref{outerthm}) and inner bound (Theorem~\ref{achievthm}) and we prove that these bounds are identical for the switch channel case. The method of our proof is similar to the method in \cite{liumaric}. According to the distribution (\ref{DistInner}), for a known auxiliary random variable $U$, auxiliary random variables $V_{1}$ and $V_{2}$ are independent, but this is not true for distribution (\ref{DistOuter}). Therefore, we first show that these distributions are identical for a switch channel case. Therefore, we show that,
\begin{equation}\label{Independance}
I(V_{1};V_{2}|U)=0
\end{equation}holds for the outer bound of switch channel case. Also, if
\begin{equation}\label{SecIndependance}
I(V_{1};V_{2}|Y_{1},U)=0,
\end{equation}holds for the outer bound of the switch channel, then
\begin{align}\label{FirstEqs}
I(V_{1};Y_{1}|V_{2},U)&=I(V_{1};Y_{1}|U)\\\label{SecondEqs}
I(V_{2};Y_{1}|V_{1},U)&=I(V_{2};Y_{1}|U)
\end{align}that is, for the switch channel case, (\ref{R1})-(\ref{R2}) in the outer bound and (\ref{thm2asliR1})-(\ref{thm2asliR2}) in the inner bound are identical.

Now, we prove that equations (\ref{Independance}) and (\ref{SecIndependance}) hold for outer bound of switch channel model. From definitions (\ref{Ui})-(\ref{V2i}) we need to show that
\begin{align}
\label{SecondEqs1}
I(W_{1};W_{2}|U_{i})&=0\\
\label{SecondEqs2}
I(W_{1};W_{2}|{U_i},Y_{1,i})&=0
\end{align}where according to (\ref{Ui}), $U_{i}={W_0},{Y_{1}^{i - 1}},Y_{2,i + 1}^n$. We first show that (\ref{SecondEqs1}) holds for switch channel. From definition (\ref{yt}) we have,
\begin{equation}\label{Y1Y222}
\{Y_{1}^{i - 1},Y_{2,i + 1}^{n}\}=\{K_{1}^{i - 1},K_{2,i + 1}^{n},S_{1}^{i-1},S_{2,i+1}^{n}\}
\end{equation}therefore,
\noindent
\makeatletter%
\if@twocolumn%
\begin{align}
&I(W_{1};W_{2}|{U_i})=I(W_{1};W_{2}|{W_0},K_{1}^{i - 1},K_{2,i + 1}^{n},S_{1}^{i-1},S_{2,i+1}^{n})\nonumber\\
&=\sum\limits_{s_1^{i - 1}} {\sum\limits_{s_{2,i + 1}^n} {P(S_1^{i - 1} = s_1^{i - 1},S_{2,i + 1}^n = s_{2,i + 1}^n)} }\nonumber\\
&\,\,\,\,\times I({W_1};{W_2}|{W_0},K_1^{i - 1},K_{2,i + 1}^n,s_1^{i - 1},s_{2,i + 1}^n)\nonumber\\
&= \sum\limits_{s_1^{i - 1}} {\sum\limits_{s_{2,i + 1}^n} {\left[ {\prod\limits_{a = 1}^{i - 1} {P({S_{1,a}} = {s_{1,a}})} \prod\limits_{b = i + 1}^n {P({S_{2,b}} = {s_{2,b}})} } \right]} }\nonumber\\
&\,\,\,\,\times I({W_1};{W_2}|{W_0},K_1^{i - 1},K_{2,i + 1}^n,s_1^{i - 1},s_{2,i + 1}^n).
\end{align}
\else
\begin{align}
&I(W_{1};W_{2}|{U_i})=I(W_{1};W_{2}|{W_0},K_{1}^{i - 1},K_{2,i + 1}^{n},S_{1}^{i-1},S_{2,i+1}^{n})\nonumber\\
&=\sum\limits_{s_1^{i - 1}} {\sum\limits_{s_{2,i + 1}^n} {P(S_1^{i - 1} = s_1^{i - 1},S_{2,i + 1}^n = s_{2,i + 1}^n)I({W_1};{W_2}|{W_0},K_1^{i - 1},K_{2,i + 1}^n,s_1^{i - 1},s_{2,i + 1}^n)} }\nonumber\\
&= \sum\limits_{s_1^{i - 1}} {\sum\limits_{s_{2,i + 1}^n} {\left[ {\prod\limits_{a = 1}^{i - 1} {P({S_{1,a}} = {s_{1,a}})} \prod\limits_{b = i + 1}^n {P({S_{2,b}} = {s_{2,b}})} } \right]I({W_1};{W_2}|{W_0},K_1^{i - 1},K_{2,i + 1}^n,s_1^{i - 1},s_{2,i + 1}^n)} }.
\end{align}
\fi
\makeatother

Now, for known $s_{t,i}$, the switch channel model (\ref{SCmodel}) shows that $k_{t,i}$ depends only on the channel input $x_{{s_{t,i}},i}$. From \cite{liumaric}, and for known switch state information $s_1^{i - 1}$ and $s_{2,i + 1}^n$ we can easily show that
\begin{align}
I({W_1};{W_2}|{W_0},K_1^{i - 1},K_{2,i + 1}^n,s_1^{i - 1},s_{2,i + 1}^n)=0.
\end{align}This proves the equality (\ref{Independance}). Proceeding the same way, we can show that equality (\ref{SecIndependance}) holds.
Hence, the equalities (\ref{FirstEqs})-(\ref{SecondEqs}) hold for the switch channel. Also, (\ref{R0}) implies (\ref{thm2asli}). Moreover, in the derived outer bound in Theorem~\ref{outerthm}, adding (\ref{R0}) to the (\ref{jameR1R2}) subject to the existing Markov chain $U-(V_1,V_2)-(Y_1,Y_2)$ gives: $R_0+R_1+R_2 \leq I({V_1},{V_2};Y_{1})-I({V_1},{V_2};Y_{2}|U)$ that is identical to (\ref{thm2asliR0R1R2}) in the derived inner bound.
The bounds on cardinality of $|\cal U|$, $|{\cal V}_1|$ and $|{\cal V}_2|$ can be derived by following the steps in \cite[Appendix]{CsiszarKorner}. This completes the proof.
\section{Proof for Theorem~\ref{outergausithm}}\label{outergausiproof}
To derive an outer bound for the Gaussian case we can follow the steps of deriving an outer bound for discrete memoryless case (i.e., Theorem~\ref{outerthm}) that are based on the basic properties of mutual information (the chain rule and positivity) and hold irrespective of the continuous or discrete nature of the channel. Therefore, by following (\ref{sumR1R222}), (\ref{outerR0}) and (\ref{outerR0dov}), it can be seen that if a rate tuple $({R_0},{R_1},{R_2})$ is achievable for the Gaussian MAWC-CM, it must hold that 
\noindent
\makeatletter%
\if@twocolumn%
\begin{align}
\label{nerkhR0sum}
&{R_0} \le \min \left\{ {\frac{1}{n}\sum\limits_{i = 1}^n {I({U_i};{Y_{1,i}}),\frac{1}{n}\sum\limits_{i = 1}^n {I({U_i};{Y_{2,i}})} } } \right\}\\\label{nerkhR1sum}
%&{R_1} \le \frac{1}{n}\sum\limits_{i = 1}^n {[I({V_{1,i}};{Y_{1,i}}|{U_i}) - I({V_{1,i}};{Y_{2,i}}|{U_i})]} \\\label{nerkhR2sum}
%&{R_2} \le \frac{1}{n}\sum\limits_{i = 1}^n {[I({V_{2,i}};{Y_{1,i}}|{U_i}) - I({V_{2,i}};{Y_{2,i}}|{U_i})]} \\\label{nerkhR1R2sum}
&{R_1} + {R_2} \le \frac{1}{n}\sum\limits_{i = 1}^n {[I({V_{1,i}},{V_{2,i}};{Y_{1,i}}|{U_i})} \nonumber\\
&\,\,\,\, - I({V_{1,i}},{V_{2,i}};{Y_{2,i}}|{U_i})] %\\\label{majookoli}
%&{R_0} + {R_1} + {R_2} \le \frac{1}{n}\sum\limits_{i = 1}^n {[I({U_i},{V_{1,i}},{V_{2,i}};{Y_{1,i}}) } \nonumber\\
%&\,\,\,\,- I({V_{1,i}},{V_{2,i}};{Y_{2,i}}|{U_i})].
\end{align}
\else
\begin{align}
\label{nerkhR0sum}
&{R_0} \le \min \left\{ {\frac{1}{n}\sum\limits_{i = 1}^n {I({U_i};{Y_{1,i}}),\frac{1}{n}\sum\limits_{i = 1}^n {I({U_i};{Y_{2,i}})} } } \right\}\\\label{nerkhR1sum}
%&{R_1} \le \frac{1}{n}\sum\limits_{i = 1}^n {[I({V_{1,i}};{Y_{1,i}}|{U_i}) - I({V_{1,i}};{Y_{2,i}}|{U_i})]} \\\label{nerkhR2sum}
%&{R_2} \le \frac{1}{n}\sum\limits_{i = 1}^n {[I({V_{2,i}};{Y_{1,i}}|{U_i}) - I({V_{2,i}};{Y_{2,i}}|{U_i})]} \\\label{nerkhR1R2sum}
&{R_1} + {R_2} \le \frac{1}{n}\sum\limits_{i = 1}^n {[I({V_{1,i}},{V_{2,i}};{Y_{1,i}}|{U_i}) - I({V_{1,i}},{V_{2,i}};{Y_{2,i}}|{U_i})]} %\\\label{majookoli}
%&{R_0} + {R_1} + {R_2} \le \frac{1}{n}\sum\limits_{i = 1}^n {[I({U_i},{V_{1,i}},{V_{2,i}};{Y_{1,i}}) } - I({V_{1,i}},{V_{2,i}};{Y_{2,i}}|{U_i})].
\end{align} 
\fi
\makeatother
It remains to upper bound (\ref{nerkhR0sum}) and (\ref{nerkhR1sum}) with terms that depend on the power constraints $P_1$ and $P_2$. We first assume that $\sigma _1^2 \le \sigma _2^2$ so that the eavesdropper's channel is stochastically degraded with respect to the main channel. We expand $\frac{1}{n}\sum\nolimits_{i = 1}^n {I({U_i};{Y_{1,i}})}$ in terms of the differential entropy as
\begin{equation}\label{basteaval}
\frac{1}{n}\sum\limits_{i = 1}^n {I({U_i};{Y_{1,i}})}  = \frac{1}{n}\sum\limits_{i = 1}^n {h({Y_{1,i}})}  - \frac{1}{n}\sum\limits_{i = 1}^n {h({Y_{1,i}}|{U_i})}
\end{equation}and we bound each sum separately. Due to channel model definition and assumptions, we have 
\begin{align}
{\mathop{\rm var}} ({Y_{1,i}}) =
%E[Y_{1,i}^2] - {\left( {E[{Y_{1,i}}]} \right)^2} = E[Y_{1,i}^2] \nonumber\\
%&= E[{({X_{1,i}} + {X_{2,i}} + {N_{1,i}})^2}]\nonumber\\
%&= E\left( {X_{1,i}^2 + X_{2,i}^2 + N_{1,i}^2 + 2{X_{1,i}}{X_{2,i}} + 2{X_{1,i}}{N_{1,i}} + 2{N_{1,i}}{X_{2,i}}} \right)\nonumber\\
%&= E[X_{1,i}^2] + E[X_{2,i}^2] + E[N_{1,i}^2] + 2E[{X_{1,i}}{X_{2,i}}] + 2E[{X_{1,i}}{N_{1,i}}] \nonumber\\
%&\,\,\,\,\,\,\,+ 2E[{N_{1,i}}{X_{2,i}}]\nonumber\\
%&= E[X_{1,i}^2] + E[X_{2,i}^2] + \sigma _{1,i}^2 + 2E[{X_{1,i}}{X_{2,i}}]\nonumber\\
E[X_{1,i}^2] + E[X_{2,i}^2] + \sigma _{1}^2 + 2{\lambda _i}
\end{align}where $\lambda _{i}=E[{X_{1,i}}{X_{2,i}}]$. So, the differential entropy of $Y_{1,i}$ is upper bounded by the entropy of a Gaussian random variable with the same variance. Hence,
\noindent
\makeatletter%
\if@twocolumn%
{\small{\begin{align}
&\frac{1}{n}\sum\limits_{i = 1}^n {h({Y_{1,i}})} \le \frac{1}{n}\sum\limits_{i = 1}^n {\frac{1}{2}\log (2\pi e(E[X_{1,i}^2] + E[X_{2,i}^2] + \sigma _{1}^2 + 2{\lambda _i}))} \nonumber\\
&\mathop  \le \limits^{(a)} %\label{Jensen1}
 \frac{1}{2}\log (2\pi e(\frac{1}{n}\sum\limits_{i = 1}^n {E[X_{1,i}^2]}  + \frac{1}{n}\sum\limits_{i = 1}^n {E[X_{2,i}^2]} + \frac{2}{n}\sum\limits_{i = 1}^n {{\lambda _i}}  + \sigma _{1}^2)),\nonumber
\end{align}}}
\else
\begin{align}
\frac{1}{n}\sum\limits_{i = 1}^n {h({Y_{1,i}})} &\le \frac{1}{n}\sum\limits_{i = 1}^n {\frac{1}{2}\log (2\pi e(E[X_{1,i}^2] + E[X_{2,i}^2] + \sigma _{1}^2 + 2{\lambda _i}))} \nonumber\\
&\mathop  \le \limits^{(a)} %\label{Jensen1}
 \frac{1}{2}\log (2\pi e(\frac{1}{n}\sum\limits_{i = 1}^n {E[X_{1,i}^2]}  + \frac{1}{n}\sum\limits_{i = 1}^n {E[X_{2,i}^2]} + \frac{2}{n}\sum\limits_{i = 1}^n {{\lambda _i}}  + \sigma _{1}^2)),\nonumber
\end{align}
\fi
\makeatother
where $(a)$ results since $x \mapsto \log (2\pi ex)$ is a concave function of $x$ as using Jensen's inequality. Also, by setting ${Q_1} \buildrel \Delta \over = \frac{1}{n}\sum\nolimits_{i = 1}^n {E[X_{1,i}^2]}$, ${Q_2} \buildrel \Delta \over = \frac{1}{n}\sum\nolimits_{i = 1}^n {E[X_{2,i}^2]}$, ${Q_3} \buildrel \Delta \over = \frac{1}{n}\sum\nolimits_{i = 1}^n {{\lambda _i}}$ and $\rho  = \frac{{{Q_3}}}{{\sqrt {{Q_1}{Q_2}} }}$ we finally obtain
\begin{align}\label{entropyY12}
\frac{1}{n}\sum\limits_{i = 1}^n {h({Y_{1,i}})}  \le \frac{1}{2}\log (2\pi e({Q_1} + {Q_2} + 2{\rho \sqrt {{Q_1}{Q_2}}} + \sigma _1^2)).
\end{align}

To bound the second sum $\frac{1}{n}\sum\nolimits_{i = 1}^n {h({Y_i}|{U_i})}$, notice that
\begin{align}\label{entropyY13}
\frac{1}{n}\sum\limits_{i = 1}^n {h({Y_{1,i}}|{U_i})}  &\le \frac{1}{n}\sum\limits_{i = 1}^n {h({Y_{1,i}})} \nonumber\\
&\le \frac{1}{2}\log (2\pi e({Q_1} + {Q_2} + 2{\rho \sqrt {{Q_1}{Q_2}}} + \sigma _1^2)).
\end{align}

Moreover, because ${U_i} \to ({V_{1,i}},{V_{2,i}}) \to ({X_{1,i}},{X_{2,i}}) \to ({Y_{1,i}},{Y_{2,i}})$ forms a Markov chain, we have
\begin{align}
\frac{1}{n}\sum\limits_{i = 1}^n {h({Y_{1,i}}|{U_i})}  &\ge \frac{1}{n}\sum\limits_{i = 1}^n {h({Y_{1,i}}|{U_i},{X_{1,i}},{X_{2,i}}) }\nonumber\\
&=  \frac{1}{n}\sum\limits_{i = 1}^n {h({Y_{1,i}}|{X_{1,i}},{X_{2,i}}) }\nonumber\\\label{entropyY1Sharty1}
&= \frac{1}{2}\log (2\pi e\sigma _1^2).
\end{align}

Since $x,y \mapsto \frac{1}{2}\log (2\pi e(x{Q_1} + y{Q_2} + 2xy{\rho \sqrt {{Q_1}{Q_2}}} + \sigma _1^2))$ is a continuous function on interval $x \in [0,1]$ and $y \in [0,1]$, a two-dimensional intermediate-value theorem ensures the existence of ${\beta _1},{\beta _2} \in [0,1]$ such that
\noindent
\makeatletter%
\if@twocolumn%
\begin{align}\label{entropyY1Sharty2}
&\frac{1}{n}\sum\limits_{i = 1}^n {h({Y_{1,i}}|{U_i})} \nonumber\\
&\,\,\,\, = \frac{1}{2}\log (2\pi e({\beta _1}{Q_1} + {\beta _2}{Q_2} + 2{\beta _1}{\beta _2}{\rho \sqrt {{Q_1}{Q_2}}} + \sigma _1^2)).
\end{align}
\else
\begin{align}\label{entropyY1Sharty2}
&\frac{1}{n}\sum\limits_{i = 1}^n {h({Y_{1,i}}|{U_i})} = \frac{1}{2}\log (2\pi e({\beta _1}{Q_1} + {\beta _2}{Q_2} + 2{\beta _1}{\beta _2}{\rho \sqrt {{Q_1}{Q_2}}} + \sigma _1^2)).
\end{align}
\fi
\makeatother

By substituting (\ref{entropyY12}) and (\ref{entropyY1Sharty2}) into (\ref{basteaval}), we obtain
\noindent
\makeatletter%
\if@twocolumn%
{\small{\begin{align}\label{Iaval}
&\frac{1}{n}\sum\limits_{i = 1}^n {I({U_i};{Y_{1,i}})} \le \frac{1}{2}\log (2\pi e({Q_1} + {Q_2} + 2{\rho \sqrt {{Q_1}{Q_2}}} + \sigma _1^2)) \nonumber\\
&\,\,\,\,\,\,- \frac{1}{2}\log (2\pi e({\beta _1}{Q_1} + {\beta _2}{Q_2} + 2{\beta _1}{\beta _2}{\rho \sqrt {{Q_1}{Q_2}}} + \sigma _1^2))\nonumber\\
&=\frac{1}{2}\log ( {1 + \frac{{(1 - {\beta _1}){Q_1} + (1 - {\beta _2}){Q_2} + 2(1 - {\beta _1}{\beta _2}){\rho \sqrt {{Q_1}{Q_2}}}}}{{{\beta _1}{Q_1} + {\beta _2}{Q_2} + 2{\beta _1}{\beta _2}{\rho \sqrt {{Q_1}{Q_2}}} + \sigma _1^2}}}).
\end{align}}}
\else
\begin{align}\label{Iaval}
\frac{1}{n}\sum\limits_{i = 1}^n {I({U_i};{Y_{1,i}})} &\le \frac{1}{2}\log (2\pi e({Q_1} + {Q_2} + 2{\rho \sqrt {{Q_1}{Q_2}}} + \sigma _1^2)) \nonumber\\
&\,\,\,\,\,\,- \frac{1}{2}\log (2\pi e({\beta _1}{Q_1} + {\beta _2}{Q_2} + 2{\beta _1}{\beta _2}{\rho \sqrt {{Q_1}{Q_2}}} + \sigma _1^2))\nonumber\\
&=\frac{1}{2}\log ( {1 + \frac{{(1 - {\beta _1}){Q_1} + (1 - {\beta _2}){Q_2} + 2(1 - {\beta _1}{\beta _2}){\rho \sqrt {{Q_1}{Q_2}}}}}{{{\beta _1}{Q_1} + {\beta _2}{Q_2} + 2{\beta _1}{\beta _2}{\rho \sqrt {{Q_1}{Q_2}}} + \sigma _1^2}}}).
\end{align}
\fi
\makeatother

Now, we need to upper bound $\frac{1}{n}\sum\nolimits_{i = 1}^n {I({U_i};{Y_{2,i}})}$. Note that we can repeat the steps leading to (\ref{entropyY12}) with $Y_{2,i}$ instead of $Y_{1,i}$ to obtain
\begin{equation}\label{entropyY21}
\frac{1}{n}\sum\limits_{i = 1}^n {h({Y_{2,i}})} \le \frac{1}{2}\log (2\pi e({Q_1} + {Q_2} + 2{\rho \sqrt {{Q_1}{Q_2}}} + \sigma _2^2)).
\end{equation}

So, we need to derive a lower bound for $\frac{1}{n}\sum\nolimits_{i = 1}^n {h({Y_{2,i}}|{U_i})}$ as a function of ${Q_1}$, ${Q_2}$, ${\beta _1}$, ${\beta _2}$ and ${\rho}$. Since we have assumed that the eavesdropper's channel is stochastically degraded with respect to the main channel, we can write ${Y_{2,i}} = {Y_{1,i}} + {N'_i}$ with ${N'_i} \sim {\cal N}(0,\sigma _2^2 - \sigma _1^2)$. Applying the Entropy Power Inequality (EPI) \cite{ElgamalKim} to the RV $Y_{2,i}$ conditioned on ${U_i} = {u_i}$, we have
\begin{align}\label{entropyY2sharty}
h({Y_{2,i}}|{U_i} = {u_i}) &= h({Y_{1,i}} + {{N'_i}}|{U_i} = {u_i})\nonumber\\
&\ge \frac{1}{2}\log \left( {{2^{2h({Y_{1,i}}|{U_i} = {u_i})}} + {2^{2h({{N'_i}}|{U_i} = {u_i})}}} \right)\nonumber\\
&= \frac{1}{2}\log \left( {{2^{2h({Y_{1,i}}|{U_i} = {u_i})}} + 2\pi e(\sigma _2^2 - \sigma _1^2)} \right).
\end{align}

Hence,
\noindent
\makeatletter%
\if@twocolumn%
\begin{align}
&\frac{1}{n}\sum\limits_{i = 1}^n {h({Y_{2,i}}|{U_i})} = \frac{1}{n}\sum\limits_{i = 1}^n {{E_{{U_i}}}[h({Y_{2,i}}|{U_i})]} \nonumber\\
&\mathop  \ge \limits^{(a)} \frac{1}{{2n}}\sum\limits_{i = 1}^n {{E_{{U_i}}}\left[ {\log \left( {{2^{2h({Y_{1,i}}|{U_i})}} + 2\pi e\left( {\sigma _2^2 - \sigma _1^2} \right)} \right)} \right]}  \nonumber\\
&\mathop  \ge \limits^{(b)} \frac{1}{{2n}}\sum\limits_{i = 1}^n {\log \left( {{2^{2{E_{{U_i}}}[h({Y_{1,i}}|{U_i})]}} + 2\pi e\left( {\sigma _2^2 - \sigma _1^2} \right)} \right)} \nonumber\\
& = \frac{1}{{2n}}\sum\limits_{i = 1}^n {\log \left( {{2^{2h({Y_{1,i}}|{U_i})}} + 2\pi e(\sigma _2^2 - \sigma _1^2)} \right)}\nonumber \\
&\mathop  \ge \limits^{(c)} \frac{1}{2}\log \left( {{2^{\frac{2}{n}\sum\nolimits_{i = 1}^n {h({Y_{1,i}}|{U_i})} }} + 2\pi e(\sigma _2^2 - \sigma _1^2)} \right)\nonumber\\\label{entropyshartiY22}
%&\mathop  = \limits^{(d)} \frac{1}{2}\log (2\pi e({\beta _1}{Q_1} + {\beta _2}{Q_2} + 2{\beta _1}{\beta _2}{\rho \sqrt {{Q_1}{Q_2}}}+ \sigma _1^2) \nonumber\\
%&\,\,\,\,\,\,\,\,\,\,\,\,\,\,\,\,\,\,\,\,\,\,\,\,\,\,\,\,\,\,\,\,\,\,\,\,\,\,\,\,\,\,\,\,\,\,\,\,\,\,\,\,\,\,\,\,\,\,\,\,\,\,\,\,\,\,\,\,\,\,\,\,\,\,\,\,\,\,\,\,\,\,\,\,\,\,\,\,\,\,\,\,\,\, + 2\pi e(\sigma _2^2 - \sigma _1^2))\nonumber\\\label{entropyshartiY22}
& \mathop  = \limits^{(d)} \frac{1}{2}\log (2\pi e({\beta _1}{Q_1} + {\beta _2}{Q_2} + 2{\beta _1}{\beta _2}{\rho \sqrt {{Q_1}{Q_2}}} + \sigma _2^2)),
\end{align}
\else
\begin{align}
\frac{1}{n}\sum\limits_{i = 1}^n {h({Y_{2,i}}|{U_i})} &= \frac{1}{n}\sum\limits_{i = 1}^n {{E_{{U_i}}}[h({Y_{2,i}}|{U_i})]} \nonumber\\
&\mathop  \ge \limits^{(a)} \frac{1}{{2n}}\sum\limits_{i = 1}^n {{E_{{U_i}}}\left[ {\log \left( {{2^{2h({Y_{1,i}}|{U_i})}} + 2\pi e\left( {\sigma _2^2 - \sigma _1^2} \right)} \right)} \right]}  \nonumber\\
&\mathop  \ge \limits^{(b)} \frac{1}{{2n}}\sum\limits_{i = 1}^n {\log \left( {{2^{2{E_{{U_i}}}[h({Y_{1,i}}|{U_i})]}} + 2\pi e\left( {\sigma _2^2 - \sigma _1^2} \right)} \right)} \nonumber\\
& = \frac{1}{{2n}}\sum\limits_{i = 1}^n {\log \left( {{2^{2h({Y_{1,i}}|{U_i})}} + 2\pi e(\sigma _2^2 - \sigma _1^2)} \right)}\nonumber \\
&\mathop  \ge \limits^{(c)} \frac{1}{2}\log \left( {{2^{\frac{2}{n}\sum\nolimits_{i = 1}^n {h({Y_{1,i}}|{U_i})} }} + 2\pi e(\sigma _2^2 - \sigma _1^2)} \right)\nonumber\\\label{entropyshartiY22}
%&\mathop  = \limits^{(d)} \frac{1}{2}\log (2\pi e({\beta _1}{Q_1} + {\beta _2}{Q_2} + 2{\beta _1}{\beta _2}{\rho \sqrt {{Q_1}{Q_2}}}+ \sigma _1^2) \nonumber\\
%&\,\,\,\,\,\,\,\,\,\,\,\,\,\,\,\,\,\,\,\,\,\,\,\,\,\,\,\,\,\,\,\,\,\,\,\,\,\,\,\,\,\,\,\,\,\,\,\,\,\,\,\,\,\,\,\,\,\,\,\,\,\,\,\,\,\,\,\,\,\,\,\,\,\,\,\,\,\,\,\,\,\,\,\,\,\,\,\,\,\,\,\,\,\, + 2\pi e(\sigma _2^2 - \sigma _1^2))\nonumber\\\label{entropyshartiY22}
& \mathop  = \limits^{(d)} \frac{1}{2}\log (2\pi e({\beta _1}{Q_1} + {\beta _2}{Q_2} + 2{\beta _1}{\beta _2}{\rho \sqrt {{Q_1}{Q_2}}} + \sigma _2^2)),
\end{align}
\fi
\makeatother
where $(a)$ follows from EPI and (\ref{entropyY2sharty}). Both $(b)$ and $(c)$ follow from the convexity of the function $x \mapsto \log ({2^x} + c)$ for $c \in \mathbb{R_+}$ and Jensen's inequality, while $(d)$ follows from (\ref{entropyY1Sharty2}). Hence,
\noindent
\makeatletter%
\if@twocolumn%
{\small{\begin{align}\label{Idovom}
&\frac{1}{n}\sum\limits_{i = 1}^n {I({U_i};{Y_{2,i}})} =  \frac{1}{n}\sum\limits_{i = 1}^n {h({Y_{2,i}}) - h({Y_{2,i}}|{U_i})} \nonumber\\
&\le \frac{1}{2}\log (2\pi e({Q_1} + {Q_2} + 2{\rho \sqrt {{Q_1}{Q_2}}} + \sigma _2^n)) \nonumber\\
&\,\,\,- \frac{1}{2}\log (2\pi e({\beta _1}{Q_1} + {\beta _2}{Q_2} + 2{\beta _1}{\beta _2}{\rho \sqrt {{Q_1}{Q_2}}} + \sigma _2^2))\nonumber\\
&= \frac{1}{2}\log ( {1 + \frac{{(1 - {\beta _1}){Q_1} + (1 - {\beta _2}){Q_2} + 2(1 - {\beta _1}{\beta _2}){\rho \sqrt {{Q_1}{Q_2}}}}}{{{\beta _1}{Q_1} + {\beta _2}{Q_2} + 2{\beta _1}{\beta _2}{\rho \sqrt {{Q_1}{Q_2}}} + \sigma _2^2}}})
\end{align}}}
\else
\begin{align}\label{Idovom}
\frac{1}{n}\sum\limits_{i = 1}^n {I({U_i};{Y_{2,i}})} &=  \frac{1}{n}\sum\limits_{i = 1}^n {h({Y_{2,i}}) - h({Y_{2,i}}|{U_i})} \nonumber\\
&\le \frac{1}{2}\log (2\pi e({Q_1} + {Q_2} + 2{\rho \sqrt {{Q_1}{Q_2}}} + \sigma _2^n)) \nonumber\\
&\,\,\,- \frac{1}{2}\log (2\pi e({\beta _1}{Q_1} + {\beta _2}{Q_2} + 2{\beta _1}{\beta _2}{\rho \sqrt {{Q_1}{Q_2}}} + \sigma _2^2))\nonumber\\
&= \frac{1}{2}\log ( {1 + \frac{{(1 - {\beta _1}){Q_1} + (1 - {\beta _2}){Q_2} + 2(1 - {\beta _1}{\beta _2}){\rho \sqrt {{Q_1}{Q_2}}}}}{{{\beta _1}{Q_1} + {\beta _2}{Q_2} + 2{\beta _1}{\beta _2}{\rho \sqrt {{Q_1}{Q_2}}} + \sigma _2^2}}})
\end{align}
\fi
\makeatother
where the inequality follows from (\ref{entropyY21}) and (\ref{entropyshartiY22}). By substituting (\ref{Iaval}) and (\ref{Idovom}) into (\ref{nerkhR0sum}), we obtain
\noindent
\makeatletter%
\if@twocolumn%
{\small{\begin{equation}\label{R0result}
{R_0} \le \min \{ \begin{array}{l}
\frac{1}{2}\log ( {1 + \frac{{(1 - {\beta _1}){Q_1} + (1 - {\beta _2}){Q_2} + 2(1 - {\beta _1}{\beta _2}){\rho \sqrt {{Q_1}{Q_2}}}}}{{{\beta _1}{Q_1} + {\beta _2}{Q_2} + 2{\beta _1}{\beta _2}{\rho \sqrt {{Q_1}{Q_2}}} + \sigma _1^2}}} ),\\
\frac{1}{2}\log ( {1 + \frac{{(1 - {\beta _1}){Q_1} + (1 - {\beta _2}){Q_2} + 2(1 - {\beta _1}{\beta _2}){\rho \sqrt {{Q_1}{Q_2}}}}}{{{\beta _1}{Q_1} + {\beta _2}{Q_2} + 2{\beta _1}{\beta _2}{\rho \sqrt {{Q_1}{Q_2}}} + \sigma _2^2}}} )
\end{array} \}.
\end{equation}}}
\else
\begin{equation}\label{R0result}
{R_0} \le \min \{ \begin{array}{l}
\frac{1}{2}\log ( {1 + \frac{{(1 - {\beta _1}){Q_1} + (1 - {\beta _2}){Q_2} + 2(1 - {\beta _1}{\beta _2}){\rho \sqrt {{Q_1}{Q_2}}}}}{{{\beta _1}{Q_1} + {\beta _2}{Q_2} + 2{\beta _1}{\beta _2}{\rho \sqrt {{Q_1}{Q_2}}} + \sigma _1^2}}} ),\\
\frac{1}{2}\log ( {1 + \frac{{(1 - {\beta _1}){Q_1} + (1 - {\beta _2}){Q_2} + 2(1 - {\beta _1}{\beta _2}){\rho \sqrt {{Q_1}{Q_2}}}}}{{{\beta _1}{Q_1} + {\beta _2}{Q_2} + 2{\beta _1}{\beta _2}{\rho \sqrt {{Q_1}{Q_2}}} + \sigma _2^2}}} )
\end{array} \}.
\end{equation}
\fi
\makeatother
Now, we derive the bound on $R_1+R_2$,
\noindent
\makeatletter%
\if@twocolumn%
\begin{align}
&\frac{1}{n}\sum\limits_{i = 1}^n {[I({V_{1,i}},{V_{2,i}};{Y_{1,i}}|{U_i}) - I({V_{1,i}},{V_{2,i}};{Y_{2,i}}|{U_i})]} \nonumber\\
&= \frac{1}{n}\sum\limits_{i = 1}^n {[I({V_{1,i}},{V_{2,i}},{X_{1,i}},{X_{2,i}};{Y_{1,i}}|{U_i})} \nonumber\\
&\,\,\,\, - I({X_{1,i}},{X_{2,i}};{Y_{1,i}}|{U_i},{V_{1,i}},{V_{2,i}})  \nonumber\\
&\,\,\,\, - I({V_{1,i}},{V_{2,i}},{X_{1,i}},{X_{2,i}};{Y_{2,i}}|{U_i}) \nonumber\\
&\,\,\,\, + I({X_{1,i}},{X_{2,i}};{Y_{2,i}}|{U_i},{V_{1,i}},{V_{2,i}})]\nonumber\\
%\end{align}
%\begin{align}
&= \frac{1}{n}\sum\limits_{i = 1}^n {[I({X_{1,i}},{X_{2,i}};{Y_{1,i}}|{U_i})} \nonumber\\
&\,\,\,\, + I({V_{1,i}},{V_{2,i}};{Y_{1,i}}|{U_i},{X_{1,i}},{X_{2,i}})\nonumber\\
&\,\,\,\, - I({X_{1,i}},{X_{2,i}};{Y_{1,i}}|{U_i},{V_{1,i}},{V_{2,i}}) - I({X_{1,i}},{X_{2,i}};{Y_{2,i}}|{U_i}) \nonumber\\
&\,\,\,\, - I({V_{1,i}},{V_{2,i}};{Y_{2,i}}|{U_i},{X_{1,i}},{X_{2,i}}) \nonumber\\
&\,\,\,\, + I({X_{1,i}},{X_{2,i}};{Y_{2,i}}|{U_i},{V_{1,i}},{V_{2,i}})]\nonumber\\
&\mathop = \limits^{(a)} \frac{1}{n}\sum\limits_{i = 1}^n {[I({X_{1,i}},{X_{2,i}};{Y_{1,i}}|{U_i})} \nonumber\\
&\,\,\,\, - I({X_{1,i}},{X_{2,i}};{Y_{1,i}}|{U_i},{V_{1,i}},{V_{2,i}}) \nonumber\\
&\,\,\,\, - I({X_{1,i}},{X_{2,i}};{Y_{2,i}}|{U_i}) + I({X_{1,i}},{X_{2,i}};{Y_{2,i}}|{U_i},{V_{1,i}},{V_{2,i}})]\nonumber\\
%\end{align}
%\begin{align}
&= \frac{1}{n}\sum\limits_{i = 1}^n {[I({X_{1,i}},{X_{2,i}};{Y_{1,i}}|{U_i}) }\nonumber\\
&\,\,\,\, - I({X_{1,i}},{X_{2,i}};{Y_{1,i}},{Y_{2,i}}|{U_i},{V_{1,i}},{V_{2,i}}) \nonumber\\
&\,\,\,\, + I({X_{1,i}},{X_{2,i}};{Y_{2,i}}|{U_i},{V_{1,i}},{V_{2,i}},{Y_{1,i}}) \nonumber\\
&\,\,\,\, - I({X_{1,i}},{X_{2,i}};{Y_{2,i}}|{U_i}) + I({X_{1,i}},{X_{2,i}};{Y_{2,i}}|{U_i},{V_{1,i}},{V_{2,i}})]\nonumber\\
&\mathop  = \limits^{(b)} \frac{1}{n}\sum\limits_{i = 1}^n {[I({X_{1,i}},{X_{2,i}};{Y_{1,i}}|{U_i}) } \nonumber\\
&\,\,\,\, - I({X_{1,i}},{X_{2,i}};{Y_{1,i}},{Y_{2,i}}|{U_i},{V_{1,i}},{V_{2,i}}) \nonumber\\
&\,\,\,\, - I({X_{1,i}},{X_{2,i}};{Y_{2,i}}|{U_i}) + I({X_{1,i}},{X_{2,i}};{Y_{2,i}}|{U_i},{V_{1,i}},{V_{2,i}})]\nonumber\\\label{tabdilnerkhR1R2}
&\mathop  \le \limits^{(c)} \frac{1}{n}\sum\limits_{i = 1}^n {[I({X_{1,i}},{X_{2,i}};{Y_{1,i}}|{U_i}) - I({X_{1,i}},{X_{2,i}};{Y_{2,i}}|{U_i})]}
\end{align}
\else
\begin{align}
&\frac{1}{n}\sum\limits_{i = 1}^n {[I({V_{1,i}},{V_{2,i}};{Y_{1,i}}|{U_i}) - I({V_{1,i}},{V_{2,i}};{Y_{2,i}}|{U_i})]} \nonumber\\
&= \frac{1}{n}\sum\limits_{i = 1}^n {[I({V_{1,i}},{V_{2,i}},{X_{1,i}},{X_{2,i}};{Y_{1,i}}|{U_i})} - I({X_{1,i}},{X_{2,i}};{Y_{1,i}}|{U_i},{V_{1,i}},{V_{2,i}})  \nonumber\\
&\,\,\,- I({V_{1,i}},{V_{2,i}},{X_{1,i}},{X_{2,i}};{Y_{2,i}}|{U_i}) + I({X_{1,i}},{X_{2,i}};{Y_{2,i}}|{U_i},{V_{1,i}},{V_{2,i}})]\nonumber\\
&= \frac{1}{n}\sum\limits_{i = 1}^n {[I({X_{1,i}},{X_{2,i}};{Y_{1,i}}|{U_i}) + I({V_{1,i}},{V_{2,i}};{Y_{1,i}}|{U_i},{X_{1,i}},{X_{2,i}})}\nonumber\\
&\,\,\,- I({X_{1,i}},{X_{2,i}};{Y_{1,i}}|{U_i},{V_{1,i}},{V_{2,i}}) - I({X_{1,i}},{X_{2,i}};{Y_{2,i}}|{U_i}) \nonumber\\
&\,\,\,- I({V_{1,i}},{V_{2,i}};{Y_{2,i}}|{U_i},{X_{1,i}},{X_{2,i}}) + I({X_{1,i}},{X_{2,i}};{Y_{2,i}}|{U_i},{V_{1,i}},{V_{2,i}})]\nonumber\\
&\mathop = \limits^{(a)} \frac{1}{n}\sum\limits_{i = 1}^n {[I({X_{1,i}},{X_{2,i}};{Y_{1,i}}|{U_i}) - I({X_{1,i}},{X_{2,i}};{Y_{1,i}}|{U_i},{V_{1,i}},{V_{2,i}})} \nonumber\\
&\,\,\,- I({X_{1,i}},{X_{2,i}};{Y_{2,i}}|{U_i}) + I({X_{1,i}},{X_{2,i}};{Y_{2,i}}|{U_i},{V_{1,i}},{V_{2,i}})]\nonumber\\
&= \frac{1}{n}\sum\limits_{i = 1}^n {[I({X_{1,i}},{X_{2,i}};{Y_{1,i}}|{U_i}) }- I({X_{1,i}},{X_{2,i}};{Y_{1,i}},{Y_{2,i}}|{U_i},{V_{1,i}},{V_{2,i}}) \nonumber\\
&\,\,\,+ I({X_{1,i}},{X_{2,i}};{Y_{2,i}}|{U_i},{V_{1,i}},{V_{2,i}},{Y_{1,i}}) - I({X_{1,i}},{X_{2,i}};{Y_{2,i}}|{U_i}) + I({X_{1,i}},{X_{2,i}};{Y_{2,i}}|{U_i},{V_{1,i}},{V_{2,i}})]\nonumber\\
&\mathop  = \limits^{(b)} \frac{1}{n}\sum\limits_{i = 1}^n {[I({X_{1,i}},{X_{2,i}};{Y_{1,i}}|{U_i}) } - I({X_{1,i}},{X_{2,i}};{Y_{1,i}},{Y_{2,i}}|{U_i},{V_{1,i}},{V_{2,i}}) \nonumber\\
&\,\,\, - I({X_{1,i}},{X_{2,i}};{Y_{2,i}}|{U_i}) + I({X_{1,i}},{X_{2,i}};{Y_{2,i}}|{U_i},{V_{1,i}},{V_{2,i}})]\nonumber\\
%\end{align}
%\begin{align}
\label{tabdilnerkhR1R2}
&\mathop  \le \limits^{(c)} \frac{1}{n}\sum\limits_{i = 1}^n {[I({X_{1,i}},{X_{2,i}};{Y_{1,i}}|{U_i}) - I({X_{1,i}},{X_{2,i}};{Y_{2,i}}|{U_i})]}
\end{align}
\fi
\makeatother
where $(a)$ follows from $I({V_{1,i}},{V_{2,i}};{Y_{1,i}}|{U_i},{X_{1,i}},{X_{2,i}}) = I({V_{1,i}},{V_{2,i}};{Y_{2,i}}|{U_i},{X_{1,i}},{X_{2,i}}) = 0$ since ${U_i} \to ({V_{1,i}},{V_{2,i}}) \to ({X_{1,i}},{X_{2,i}}) \to ({Y_{1,i}},{Y_{2,i}})$ forms a Markov chain, $(b)$ follows from $I({X_{1,i}},{X_{2,i}};{Y_{2,i}}|{U_i},{V_{1,i}},{V_{2,i}},{Y_{1,i}}) = 0$ since $Y_{2,i}$ is stochastically degraded with respect to ${Y_{1,i}}$, and $(c)$ follows from $I({X_{1,i}},{X_{2,i}};{Y_{2,i}}|{U_i},{V_{1,i}},{V_{2,i}}) \le I({X_{1,i}},{X_{2,i}};{Y_{1,i}},{Y_{2,i}}|{U_i},{V_{1,i}},{V_{2,i}})$. Next, we use (\ref{entropyY1Sharty2}) and (\ref{entropyshartiY22}) to substitute ${Q_1},\,\,{Q_2},\,\,{\rho},\,\,{\beta _1}$ and $\beta _2$ as follows:
%%%%%%%%%%%%%%%%%%%%%%%%%%%%%%%%%%%%%%%%%%%ta inja%%%%%%%%%%%%%%%%%%%%%%%%%%%%%%%%%%%%
\noindent
\makeatletter%
\if@twocolumn%
\begin{align}
& \frac{1}{n}\sum\limits_{i = 1}^n {[I({V_{1,i}},{V_{2,i}};{Y_{1,i}}|{U_i}) - I({V_{1,i}},{V_{2,i}};{Y_{2,i}}|{U_i})]} \nonumber\\
&\mathop  \le \limits^{(a)} \frac{1}{n}\sum\limits_{i = 1}^n {[I({X_{1,i}},{X_{2,i}};{Y_{1,i}}|{U_i}) - I({X_{1,i}},{X_{2,i}};{Y_{2,i}}|{U_i})]} \nonumber%\\
\end{align}
\begin{align}
& = \frac{1}{n}\sum\limits_{i = 1}^n {[h({Y_{1,i}}|{U_i}) - h({Y_{1,i}}|{U_i},{X_{1,i}},{X_{2,i}})}\nonumber\\
&\,\,\,\, - h({Y_{2,i}}|{U_i}) + h({Y_{2,i}}|{U_i},{X_{1,i}},{X_{2,i}})] \nonumber\\
%\end{align}
&\mathop  \le \limits^{(b)} \frac{1}{2}\log \left( {1 + \frac{{{\beta _1}{Q_1} + {\beta _2}{Q_2} + 2{\beta _1}{\beta _2}{\rho \sqrt {{Q_1}{Q_2}}}}}{{\sigma _1^2}}} \right)  \nonumber\\\label{R1R2result}
&\,\,\,- \frac{1}{2}\log \left( {1 + \frac{{{\beta _1}{Q_1} + {\beta _2}{Q_2} + 2{\beta _1}{\beta _2}{\rho \sqrt {{Q_1}{Q_2}}}}}{{\sigma _2^2}}} \right)
\end{align}
\else
\begin{align}
& \frac{1}{n}\sum\limits_{i = 1}^n {[I({V_{1,i}},{V_{2,i}};{Y_{1,i}}|{U_i}) - I({V_{1,i}},{V_{2,i}};{Y_{2,i}}|{U_i})]} \nonumber\\
&\mathop  \le \limits^{(a)} \frac{1}{n}\sum\limits_{i = 1}^n {[I({X_{1,i}},{X_{2,i}};{Y_{1,i}}|{U_i}) - I({X_{1,i}},{X_{2,i}};{Y_{2,i}}|{U_i})]} \nonumber\\
& = \frac{1}{n}\sum\limits_{i = 1}^n {[h({Y_{1,i}}|{U_i}) - h({Y_{1,i}}|{U_i},{X_{1,i}},{X_{2,i}})}- h({Y_{2,i}}|{U_i}) + h({Y_{2,i}}|{U_i},{X_{1,i}},{X_{2,i}})] \nonumber%\\
\end{align}
\begin{align}
%& \frac{1}{n}\sum\limits_{i = 1}^n {[I({V_{1,i}},{V_{2,i}};{Y_{1,i}}|{U_i}) - I({V_{1,i}},{V_{2,i}};{Y_{2,i}}|{U_i})]} \nonumber\\
%&\le \frac{1}{2}\log 2\pi e\left( {{\beta _1}{Q_1} + {\beta _2}{Q_2} + 2{\beta _1}{\beta _2}{Q_{_3}} + \sigma _1^2} \right) - \frac{1}{2}\log \left( {2\pi e\sigma _1^2} \right)\nonumber\\
%&\,\,\, - \frac{1}{2}\log 2\pi e\left( {{\beta _1}{Q_1} + {\beta _2}{Q_2} + 2{\beta _1}{\beta _2}{Q_{_3}} + \sigma _2^2} \right) + \frac{1}{2}\log \left( {2\pi e\sigma _2^2} \right)\nonumber\\
&\mathop  \le \limits^{(b)} \frac{1}{2}\log \left( {1 + \frac{{{\beta _1}{Q_1} + {\beta _2}{Q_2} + 2{\beta _1}{\beta _2}{\rho \sqrt {{Q_1}{Q_2}}}}}{{\sigma _1^2}}} \right)  \nonumber\\\label{R1R2result}
&\,\,\,- \frac{1}{2}\log \left( {1 + \frac{{{\beta _1}{Q_1} + {\beta _2}{Q_2} + 2{\beta _1}{\beta _2}{\rho \sqrt {{Q_1}{Q_2}}}}}{{\sigma _2^2}}} \right)
\end{align}
\fi
\makeatother
where $(a)$ is due to (\ref{tabdilnerkhR1R2}), and $(b)$ results from (\ref{entropyY1Sharty1}), (\ref{entropyY1Sharty2}) and (\ref{entropyshartiY22}).

If $\sigma _1^2 \ge \sigma _2^2$, then the main channel is stochastically degraded with respect to the eavesdropper's channel and ${R_1} = {R_2} = 0$ by virtue of \cite[Proposition 3.4]{BlochBarros}. By swapping the roles of $Y_{1,i}$ and $Y_{2,i}$ in the proof, it can be verified that (\ref{R0result}) still holds. We combine the two cases $\sigma _1^2 \le \sigma _2^2$ and $\sigma _1^2 \ge \sigma _2^2$ by writing (\ref{outer}). Notice that (\ref{R0result}) and (\ref{R1R2result}) are increasing functions of $Q_1$ and $Q_2$. So, by defining $Q_j=(1/n)\sum\nolimits_{i = 1}^n {E[X_{j,i}^2]} \le {P_j},\,\,\,j = 1,2$ the inequalities in Theorem~\ref{outergausithm} hold.
This completes the proof.
\vspace{-3.5mm}
\bibliographystyle{IEEEtran}
\bibliography{IEEEexample,mybibfile}
\vspace{-3.5mm}
\end{document}